\documentclass[a4paper]{revtex4}
\usepackage[top=1in, bottom=1in, left=1in, right=1in]{geometry}
\usepackage{amsmath}
\usepackage{amssymb}
\usepackage{amsfonts}
\usepackage{graphicx,bm}
\usepackage{dcolumn}
\usepackage{epsfig}

\def \lleq {\lower0.9ex\hbox{ $\buildrel < \over \sim$} ~}
\def \ggeq {\lower0.9ex\hbox{ $\buildrel > \over \sim$} ~}

\def \om   {\Omega_m}

\def \omm  {\Omega_{0 {\rm m}}}

\def \beq  {\begin{equation}}
\def \eeq  {\end{equation}}
\def \ber  {\begin{eqnarray}}
\def \eer  {\end{eqnarray}}

\newcommand{\be}{\begin{equation}}
\newcommand{\ee}{\end{equation}}
\newcommand{\ba}{\begin{eqnarray}}
\newcommand{\ea}{\end{eqnarray}}
\newcommand{\bea}{\begin{eqnarray*}}
\newcommand{\eea}{\end{eqnarray*}}

\begin{document}
\title{Light mass galileon and late time acceleration of the Universe}

\author{ R. Myrzakulov$^1$\footnote{E-mail address: rmyrzakulov@gmail.com},
M. Shahalam$^2$\footnote{E-mail address: mohdshahamu@gmail.com}}

\affiliation{$^1$Department of General and Theoretical Physics, Eurasian National University, Astana, Kazakhstan\\
$^2$Center For Theoretical Physics, Jamia Millia Islamia, New Delhi, India}

\begin{abstract}
We study Galileon scalar field model by considering the lowest order Galileon term in the  lagrangian , $(\partial_{\mu} \phi)^2 \Box\phi$ by invoking a field potential. We use Statefinder hierarchy to distinguish the light mass galileon models with different potentials amongst themselves and from the $\Lambda$CDM behaviour. The $Om$ diagnostic is applied to cosmological dynamics and  observational constraints on the model parameters are studied using SN+Hubble+BAO data.
\end{abstract}

\date{\today}

\keywords {dark energy theory, modified gravity }
\maketitle


\section{Introduction}
\label{intro}
The late time cosmic acceleration is supported by the cosmological observations  directly \cite{obs1} and indirectly \cite{cmb,wmap}. Dark energy might be responsible for driving the cosmic acceleration of 
 Universe \cite{samiDE}. Cosmological constant is one of the simplest candidate of dark energy however it is plagued by the serious problems such as fine tuning and cosmic coincidence \cite{derev}. To understand the nature of dark energy, this is important to understand whether it is cosmological constant  or it has dynamics. The scalar field models of dark energy \cite{quint} were introduced to give a dynamical solution to the cosmological constant problem.

Recently a class of dynamical dark energy models based on the large scale modification of gravity have been proposed to describe the late time acceleration of Universe and Galileon gravity is one of them. The action of Galileon field (in absence of potential) is invariant under Galilean shift symmetry 
 $\phi(x) \rightarrow \phi(x)+b_{\mu} x^{\mu} + c$ in the Minkowski background, where $b_{\mu}$ and $c$ are the constant four vector and scalar respectively. Nicolis {\it et al.} \cite{galileon} considered five field Lagrangians ${\cal L}_i$ ($i=1,\cdots, 5$) in four dimensional flat space time. ${\cal L}_1$ is linear, ${\cal L}_2$ represents the standard kinetic term, ${\cal L}_3=(\partial_{\mu} \phi)^2 \Box\phi$ is the Vainshtein term  which has three galileon fields, and this term is associated to the decoupling limit of Dvali, Gabadadze, and Porrati (DGP) model \cite{dgp}. ${\cal L}_4$ and ${\cal L}_5$ accommodate higher order non linear derivative terms with four and five $\phi's$ respectively. Cosmological dynamics in flat FRW Universe with these terms has been investigated in reference \cite{galileon2}.
 
At least one of the higher order Galileon Lagrangian is needed to obtain a stable de sitter solution \cite{galileon3}. In this paper we focus on  ${\cal L}_3$ but add a general potential term to galileon field. We use Statefinder hierarchy to differentiate the light mass galileon models with different potentials amongst themselves and from the $\Lambda$CDM behaviour. The $Om$ diagnostic is applied to cosmological dynamics and  observational constraints on the model parameters are studied using SN+Hubble+BAO data jointly. The paper is organized as follows. The equations of motion of light mass galileon are presented in section \ref{LMG}. In section \ref{state}, the statefinder hierarchy and late time cosmological evolution is studied. The $Om$ diagnostic is discussed in section \ref{Om}. We investigate the constraints on the model parameters by applying latest observational data in section \ref{obs}. 
\section{Equations of motion}
\label{LMG}
Let us consider the action for Galileon field keeping upto the third order term in the lagrangian with a field potential  $V(\phi)$ in the action.
\begin{equation}
S=\int d^4x\sqrt{-g}\Bigl [\frac{M^2_{\rm{pl}}}{2} R -  \frac{1}{2}(\nabla \phi)^2\Bigl(1+\frac{\beta}{M^3}\Box \phi\Bigr) - V(\phi) \Bigr] + \mathcal{S}_m.
\label{eq:action}
\end{equation}
Here, $M_{\rm{pl}}^2=1/8\pi G$ is the reduced Planck mass. $\beta$ is a  dimensionless constant.  ${\cal S}_{m}$ designates matter action. $M$ is a constant of mass dimension one;  we fix $M=M_{\rm{pl}}$.

In a homogenous isotropic flat FRW Universe, the equations of motion are obtained by varying the action (eq (\ref{eq:action})) with respect to metric tensor $g_{\mu \nu}$ and scalar field $\phi$,
\begin{align}
3M_{\rm{pl}}^2H^2 &=\rho_m+\frac{\dot{\phi}^2}{2}\Bigl(1-6\frac{\beta}{M_{\rm{pl}}^{3}} H\dot{\phi}\Bigr)+V{(\phi)}\,,\\
M_{pl}^2(2\dot H + 3H^2)&=-\frac{\dot{\phi}^2}{2}\Bigl(1+2\frac{\beta}{M_{\rm{pl}}^{3}}\ddot{\phi}\Bigr)+V(\phi),
\end{align}

\begin{align}
3H\dot{\phi}+\ddot{\phi}-3\frac{\beta}{M_{\rm{pl}}^{3}} \dot{\phi}\Bigl(3H^2\dot{\phi}+\dot{H}\dot{\phi}+2H\ddot{\phi}\Bigr)+ V'(\phi)=0,
\end{align}
The above equations are augmented by the matter conservation equation,
\begin{equation}
\dot\rho_m+3H\rho_m=0.
\end{equation}

We introduce the following dimensionless quantities

\begin{align}
x&=\frac{\dot{\phi}}{\sqrt{6}H M_{\rm{pl}}}\,,\quad y=\frac{\sqrt{V}}{\sqrt{3} H M_{\rm{pl}}}\\
\epsilon &=-6\frac{\beta}{M_{\rm{pl}}^3}H\dot \phi\,, \quad \lambda=-M_{\rm{pl}}\frac{V'}{V}
\end{align}
to form an autonomous system of evolution equations:
\begin{align}
x'&=x\Bigl(\frac{\ddot{\phi}}{H\dot{\phi}}-\frac{\dot H}{H^2}\Bigr),\\
y'&=-y \Bigl(\sqrt{\frac{3}{2}}\lambda x+\frac{\dot H}{H^2}\Bigr),\\
\epsilon' &=\epsilon \Bigl(\frac{\ddot{\phi}}{H\dot{\phi}}+\frac{\dot H}{H^2}\Bigr),\\
\lambda' &=\sqrt{6}x\lambda^2(1-\Gamma),
\end{align}
where prime ($'$) denotes derivative with respect to $\ln a$, $\Gamma=\frac{VV_{,\phi\phi}}{V_{,\phi}^2}$ and
\begin{figure}
\begin{center}
\begin{tabular}{c c }
{\includegraphics[width=2.3in,height=2.1in,angle=0]{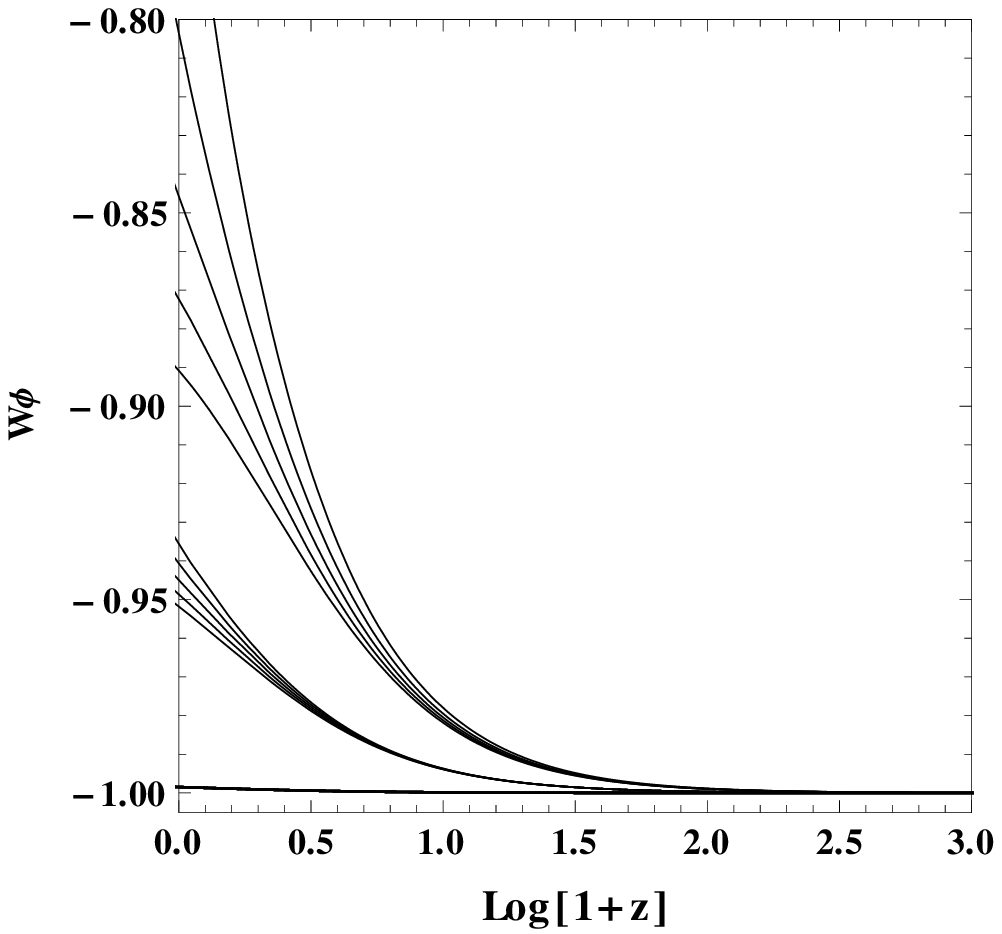}}&
{\includegraphics[width=2.3in,height=2.1in,angle=0]{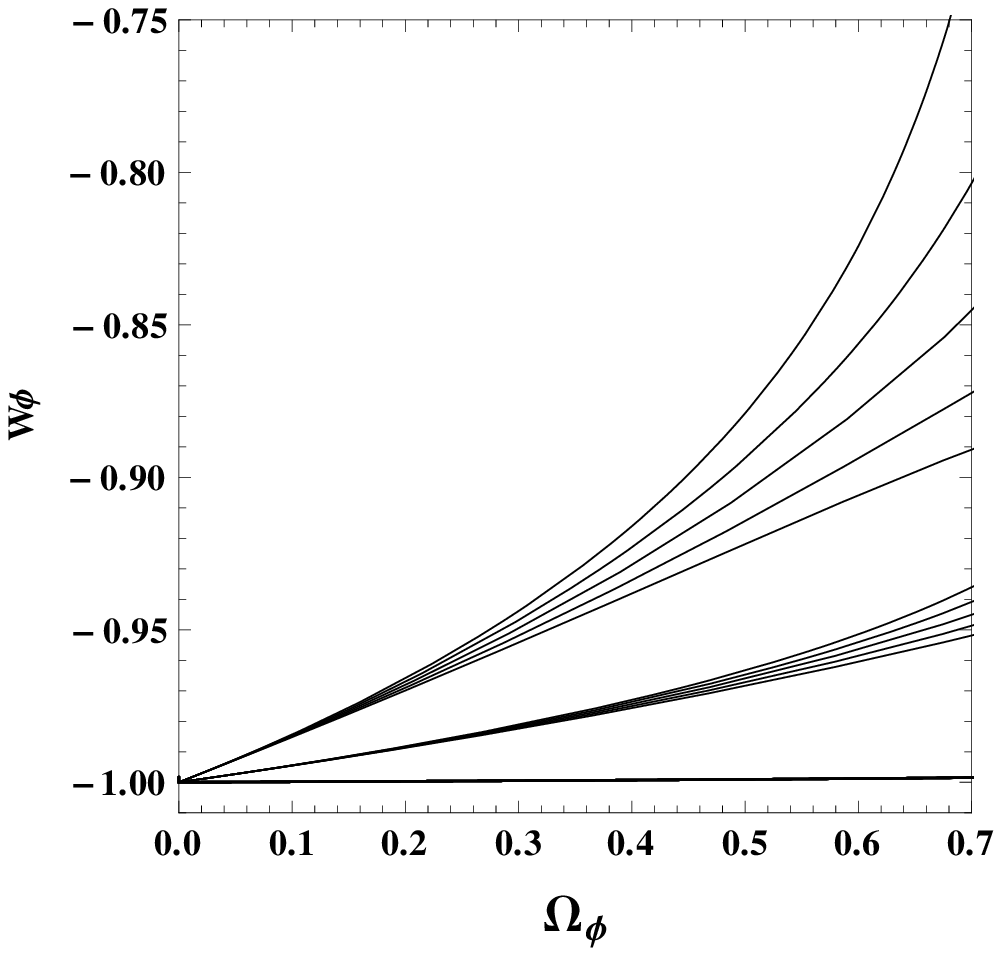}}
\end{tabular}
\caption{ This figure shows the evolution of w$_\phi$ versus the redshift z and $\Omega_\phi$. There are three sets for $\lambda_i = 0.1, 0.6, 1$ from bottom to top and $V(\phi) \sim \phi, \phi^2, e^{\phi}, \phi^{-2}, \phi^{-1}$ from top to bottom in every set.}
\label{figwpi}
\end{center}
\end{figure}
\begin{align}
\frac{\dot H}{H^2}&=\frac{2(1+\epsilon)(-3+3y^2)-3x^2(2+4\epsilon+\epsilon^2)+\sqrt{6}x\epsilon y^2\lambda}{4+4\epsilon+x^2\epsilon^2},\\
\nonumber\\
\frac{\ddot{\phi}}{H\dot{\phi}}&=\frac{3x^3\epsilon-x\Bigl(12+\epsilon (3+3y^2)\Bigr)+2\sqrt{6}y^2\lambda}{x(4+4\epsilon+x^2\epsilon^2)},
\end{align}
The equation of state for the field $\phi$ is given as,
\begin{align}
w_{eff}&= -1 -\frac{2\dot H}{3H^2},\\
w_{\phi}&= \frac{w_{eff}-w_m \Omega_m}{1-\Omega_m},
\end{align}
where $w_m = 0$ for standard dust matter. We evolve the system from $z \approx 1000$ (decoupling era) till any redshift we wish. We assume the $\phi$ field was frozen initially due to large hubble damping. This is alike to the thawing class of models \cite{scherrer}. We choose different potentials for which $\Gamma=\frac{VV_{,\phi\phi}}{V_{,\phi}^2}$= constant. 
\begin{figure}
\begin{center}
\begin{tabular}{c c }
{\includegraphics[width=2.3in,height=2.1in,angle=0]{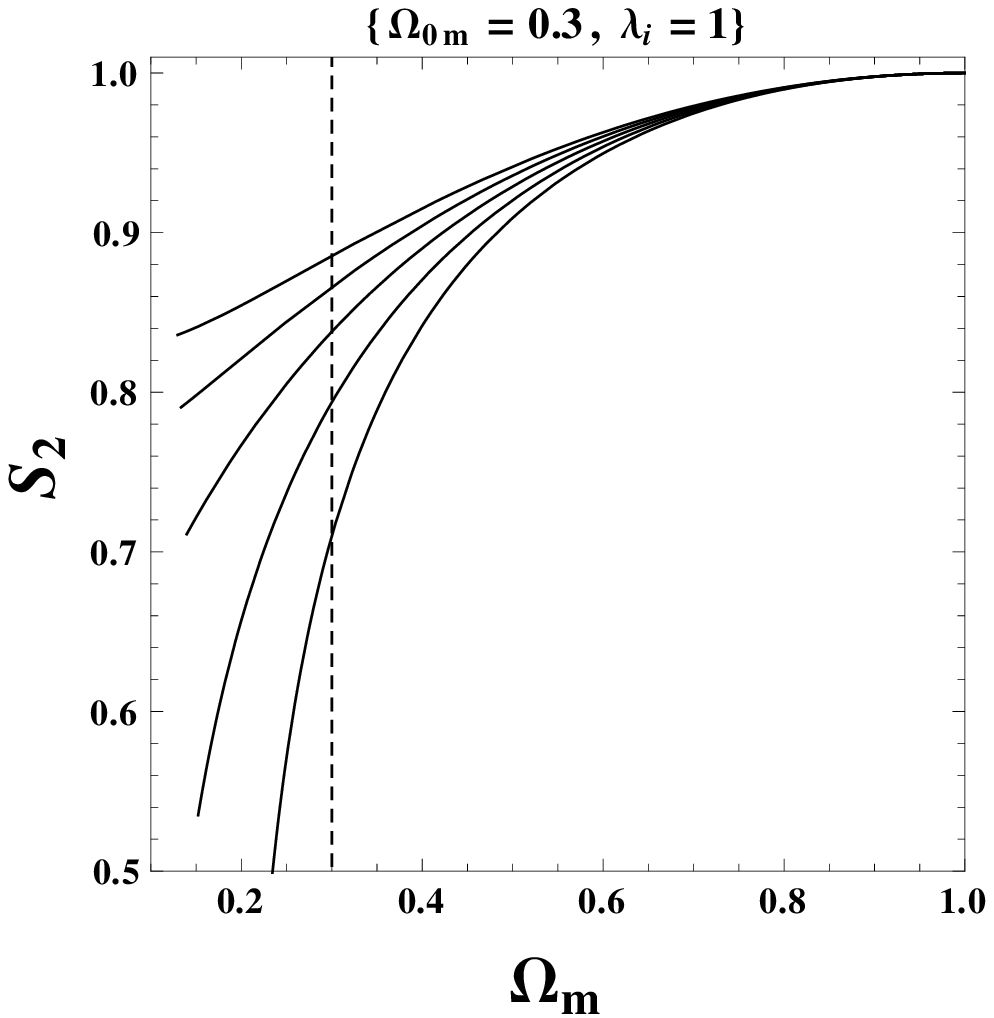}}&
{\includegraphics[width=2.3in,height=2.1in,angle=0]{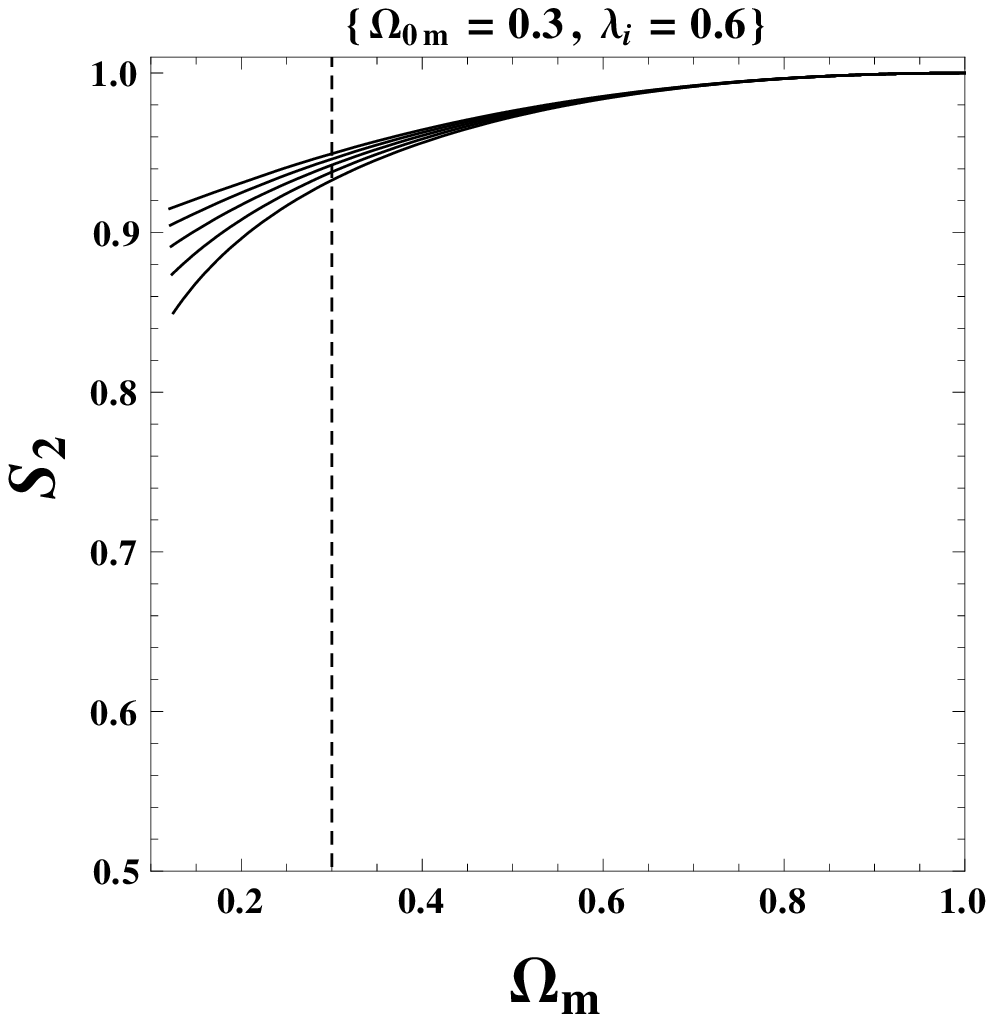}}
\\
{\includegraphics[width=2.3in,height=2.1in,angle=0]{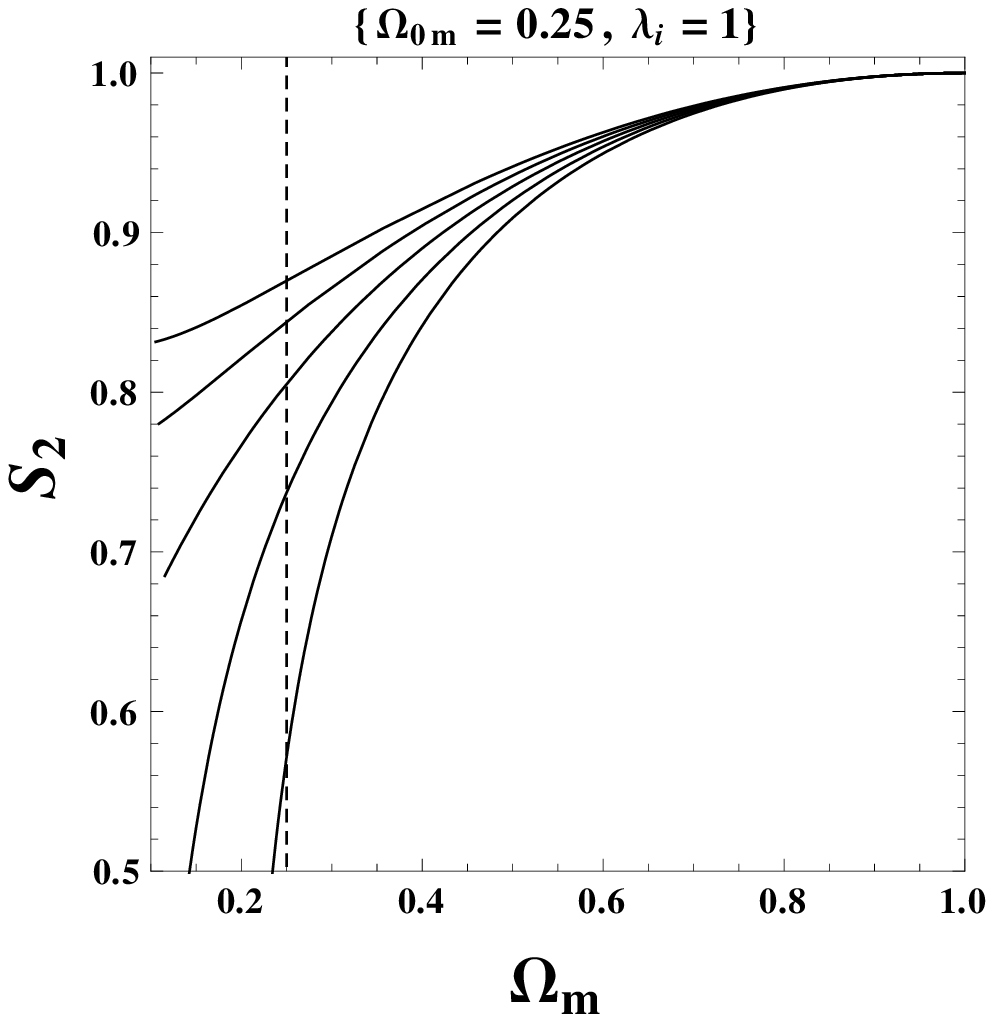}}&
{\includegraphics[width=2.3in,height=2.1in,angle=0]{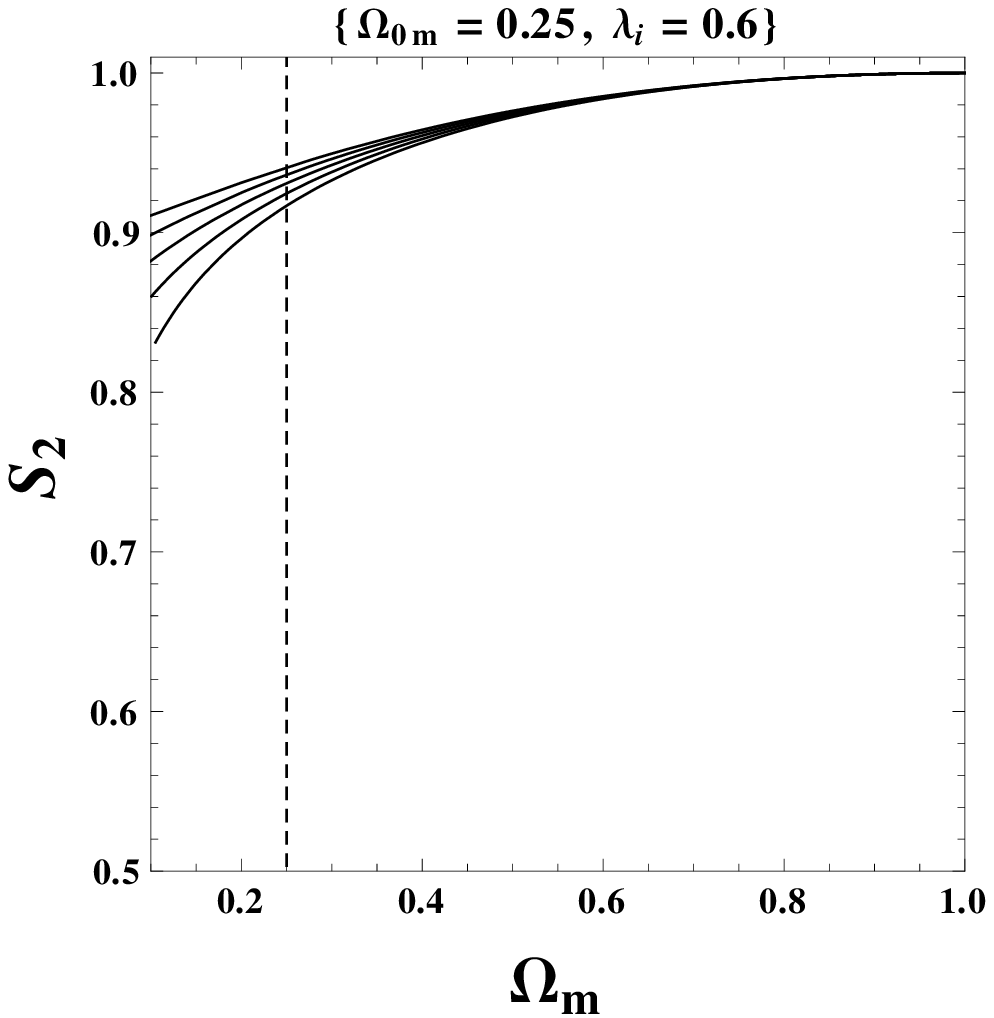}}
\end{tabular}
\caption{ This figure shows the evolution of $S_{2}$ versus $\Omega_{m}$ for different potentials $V(\phi) \sim \phi, \phi^2, e^{\phi}, \phi^{-2}, \phi^{-1}$ from bottom to top and for different values of $\lambda_{i}$ and $\Omega_{0m}$. The vertical dashed line shows the present epoch (z = 0).}
\label{figs2}
\end{center}
\end{figure}
\begin{figure}
\begin{center}
\begin{tabular}{c c }
{\includegraphics[width=2.3in,height=2.1in,angle=0]{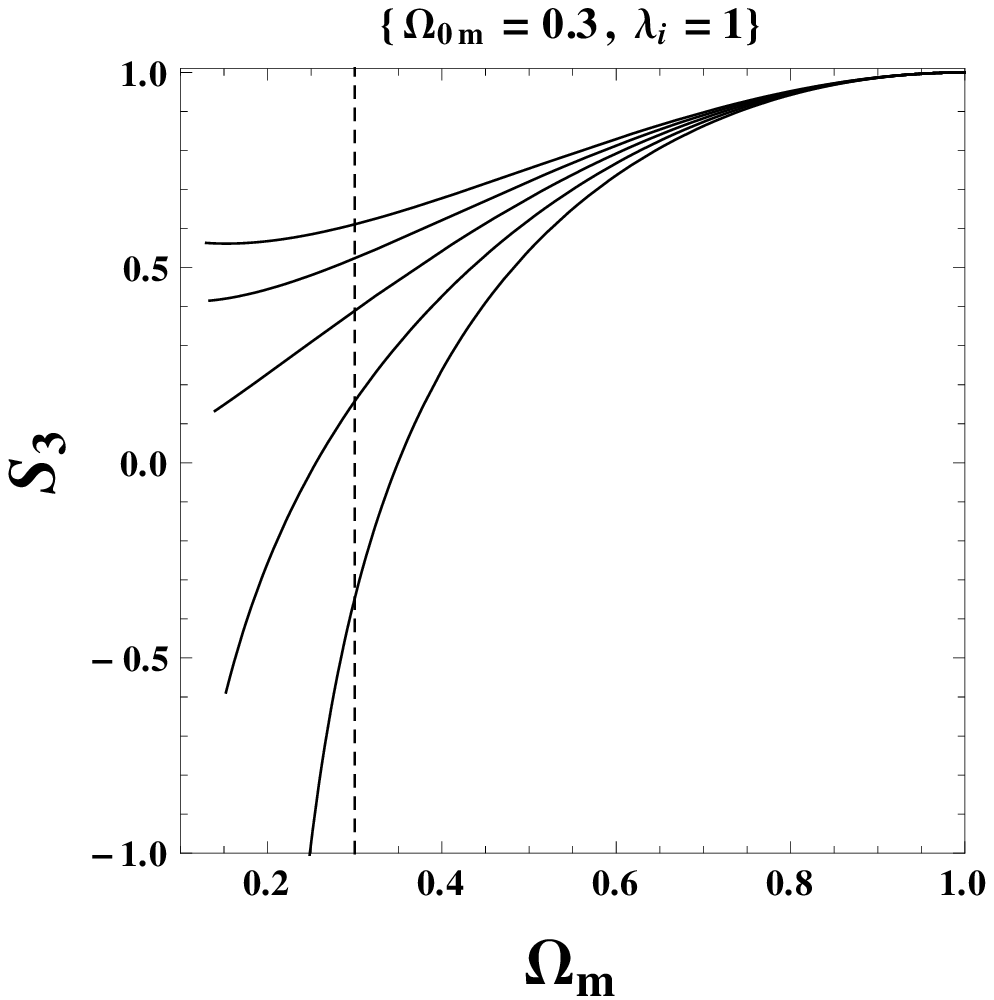}}&
{\includegraphics[width=2.3in,height=2.1in,angle=0]{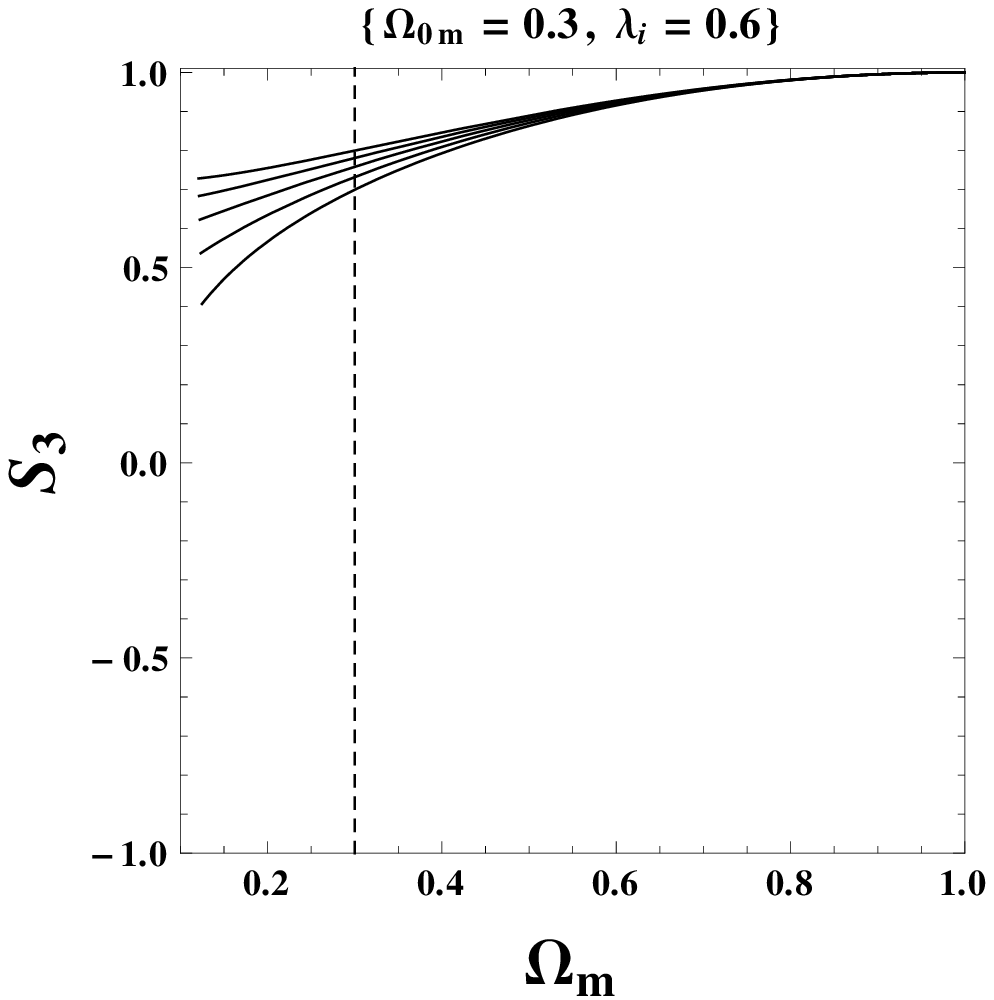}}
\\
{\includegraphics[width=2.3in,height=2.1in,angle=0]{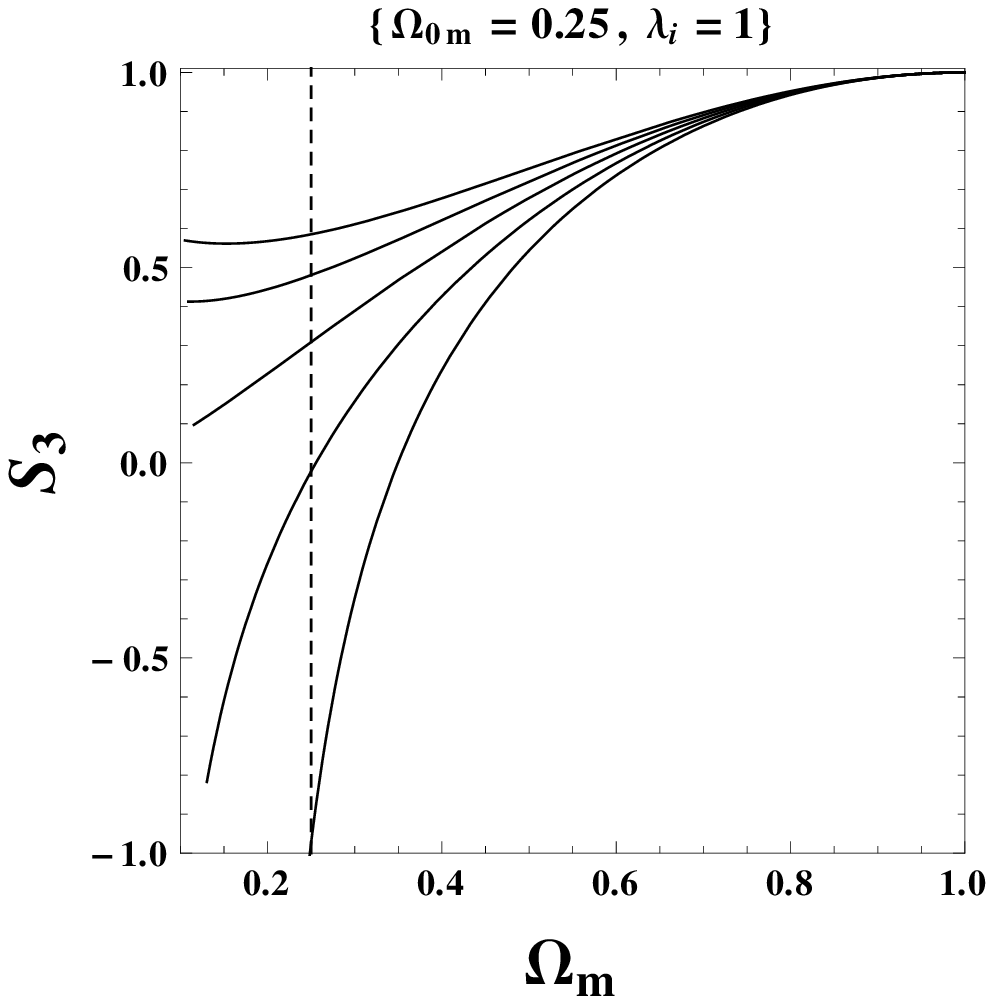}}&
{\includegraphics[width=2.3in,height=2.1in,angle=0]{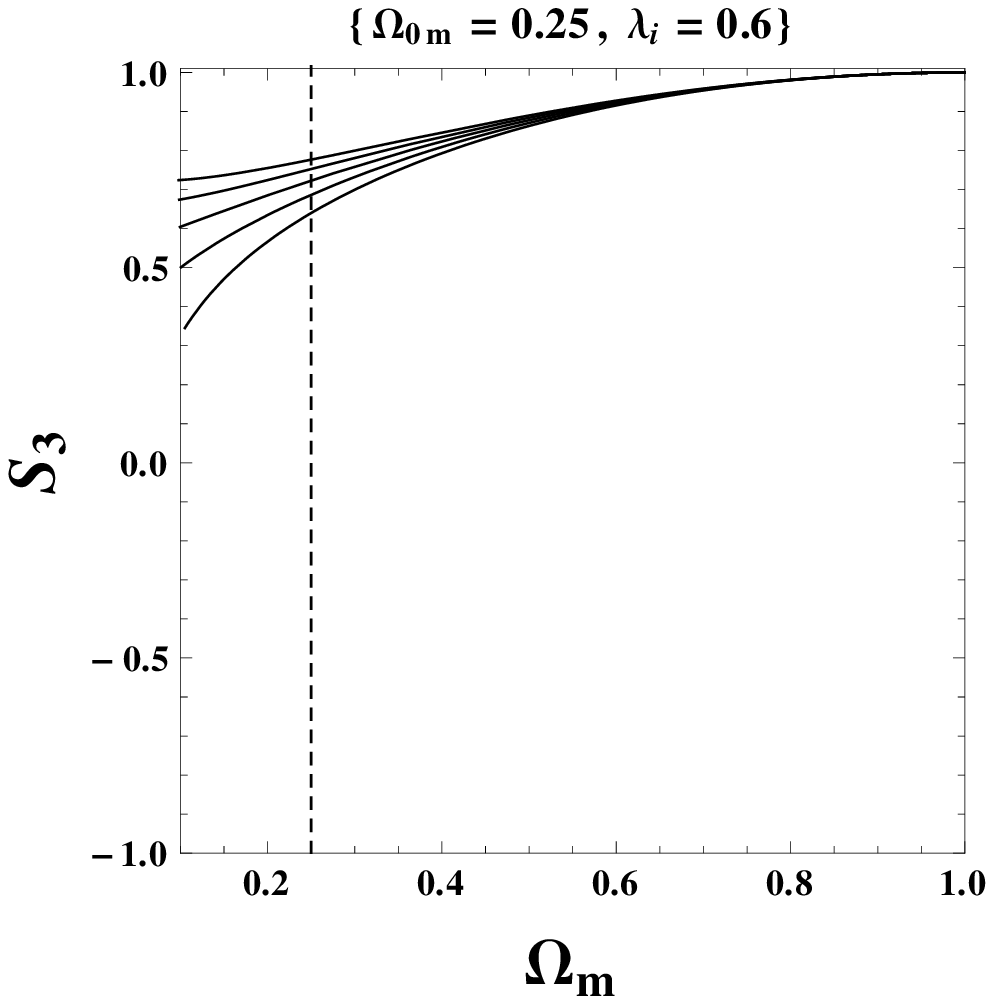}}
\end{tabular}
\caption{ Same as figure \ref{figs2} but for $S_{3}$.}
\label{figs3}
\end{center}
\end{figure}
\section{The Statefinder hierarchy and late time cosmological evolution}
\label{state}
Consider the  Taylor expansion of the scale factor around the
present era ($t=t_0$) as:
\begin{align} 
\frac{a(t)}{a(t_{0})} = 1 + \sum^{\infty}_{n=1}
\frac{\alpha_{n}(t_{0})}{n!}\left[H_{0}(t-t_{0})\right]^n,
\label{scale} 
\end{align}
where,
\begin{align}
\alpha_n=\frac{d^{n}a}{dt^{n}} / {(aH^n)},
\label{alpha}
\end{align} 
It is easy to see that $-\alpha_2 = q$ is the deceleration parameter; $\alpha_3$ and $\alpha_4$ is associated to the Statefinder $r$ and Snap $s$ respectively and so on. These parameters in terms of hubble parameter can be written as,
\begin{align}
\label{eq:a2}
\alpha_2 &= \frac{\ddot a}{aH^2} \equiv \frac{\dot H}{H^2} + 1\\
\alpha_3 &= \frac{\dddot a}{a H^3} \equiv \frac{\ddot H}{H^3} +  3\frac{\dot H}{H^2} + 1\\
\alpha_4 &=
\frac{\ddddot a}{a H^4}  \equiv 1 + \frac{\dddot H}{H^4} +
4\frac{\ddot H}{H^3} + 3\frac{\dot H^2}{H^4} + 6\frac{\dot H}{H^2}~~~~\mbox{and~so~on}
\end{align}
Using equations (\ref{scale}) and (\ref{alpha}) Arabsalmani {\it et al.} \cite{arab} define Statefinder hierarchy as:
\begin{eqnarray}
&&S_2 := \alpha_2 + \frac{3}{2}\om,\\
&&S_3  := \alpha_3,\\
&&S_4 := \alpha_4+ \frac{3^2}{2}\om,\\
&&S_5 := \alpha_5 -3\om- \frac{3^3}{2}\om^2,\\
&&S_6 := \alpha_6 + \frac{3^3}{2}\om+3^4 \om^2 +\frac{3^4}{4}\om^3 ~~~~\mbox{and~so~on}
\label{state0}
\end{eqnarray}
\begin{figure}
\begin{center}
\begin{tabular}{c c }
{\includegraphics[width=2.1in,height=2.1in,angle=0]{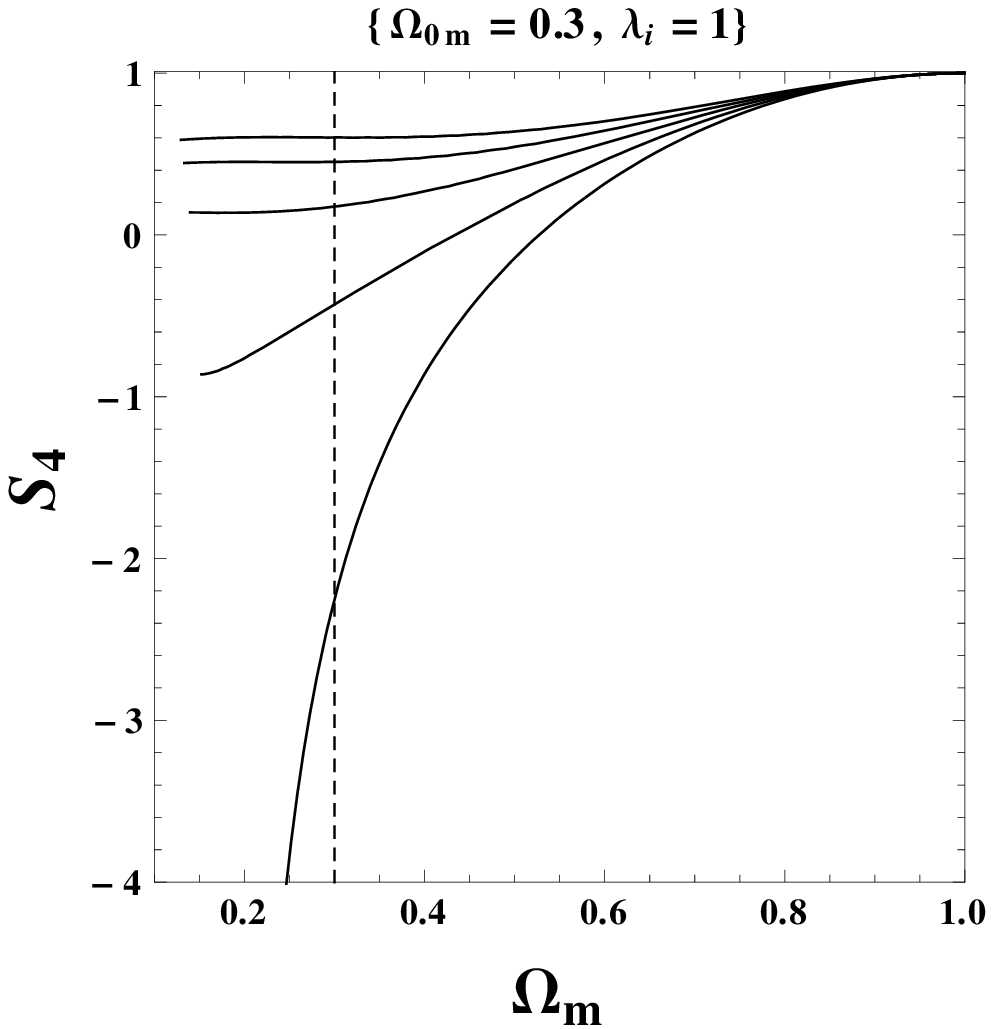}}&
{\includegraphics[width=2.1in,height=2.1in,angle=0]{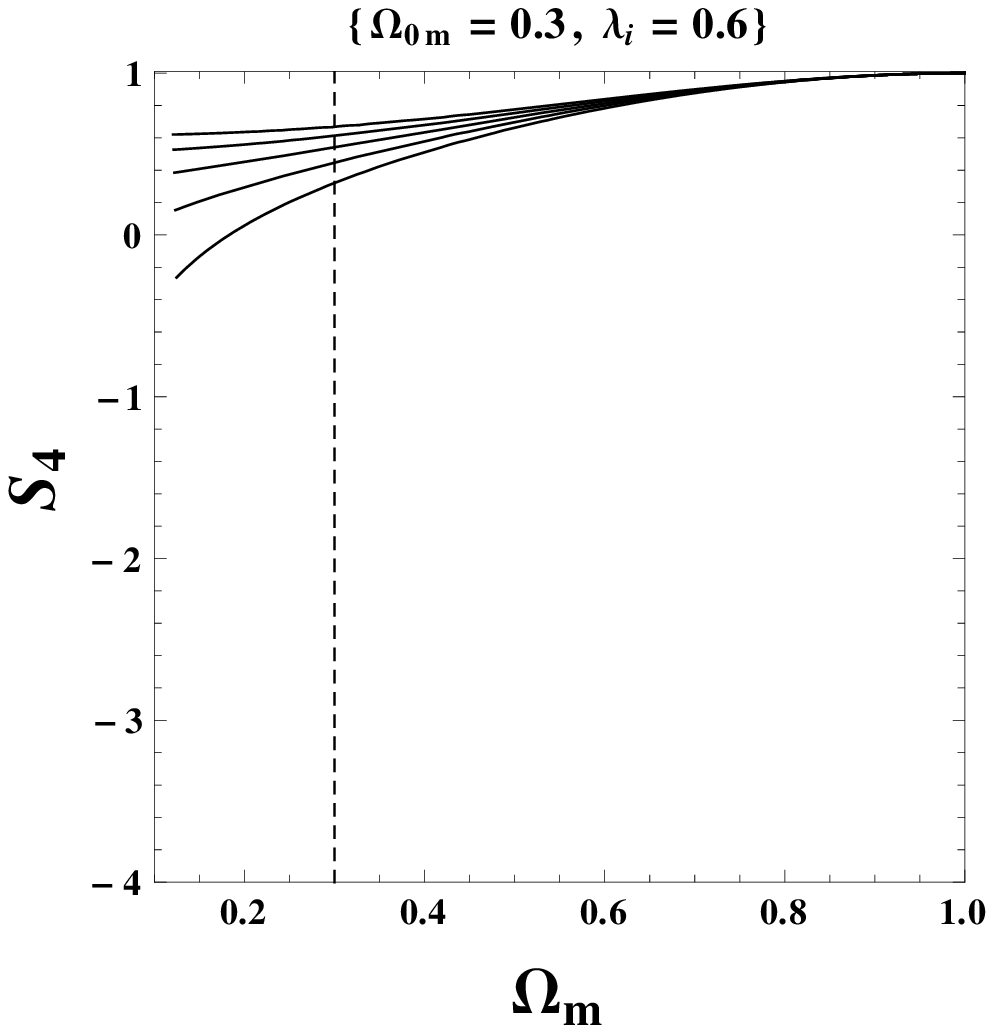}}
\\
{\includegraphics[width=2.1in,height=2.1in,angle=0]{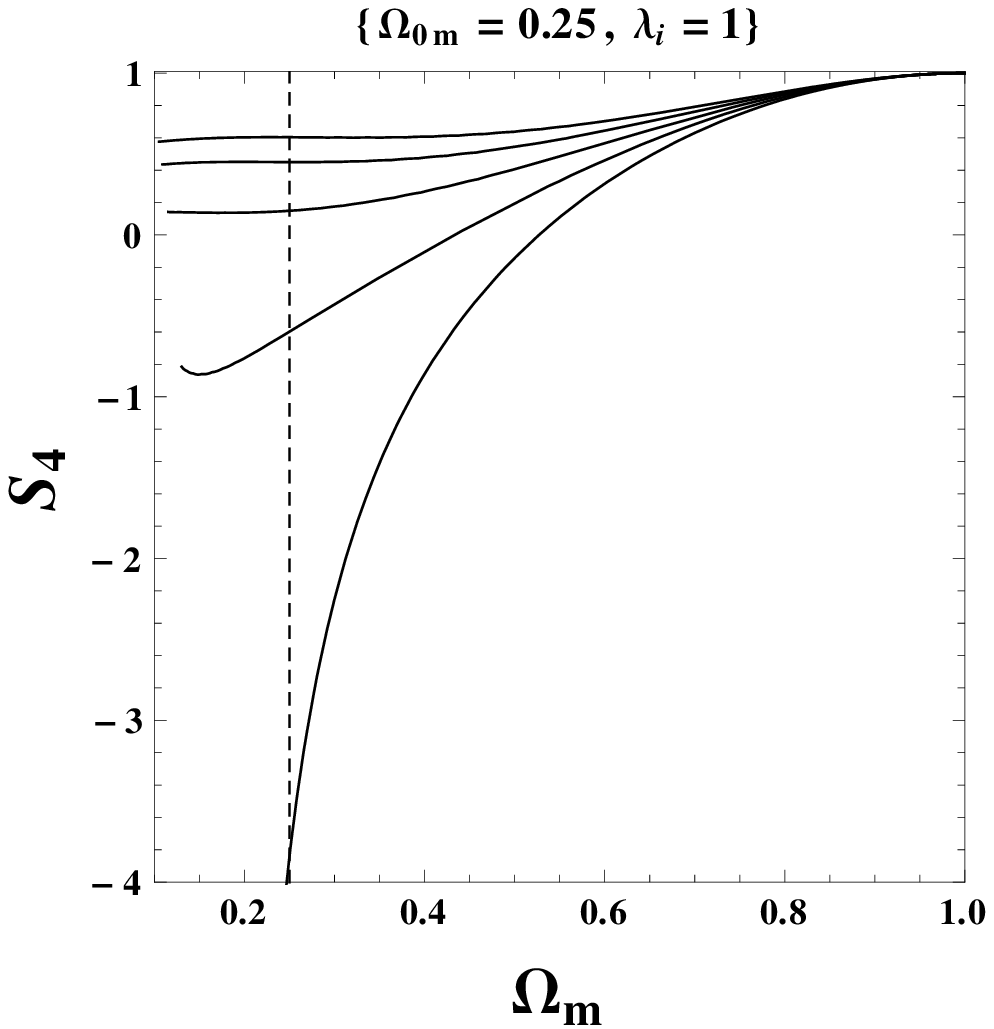}}&
{\includegraphics[width=2.1in,height=2.1in,angle=0]{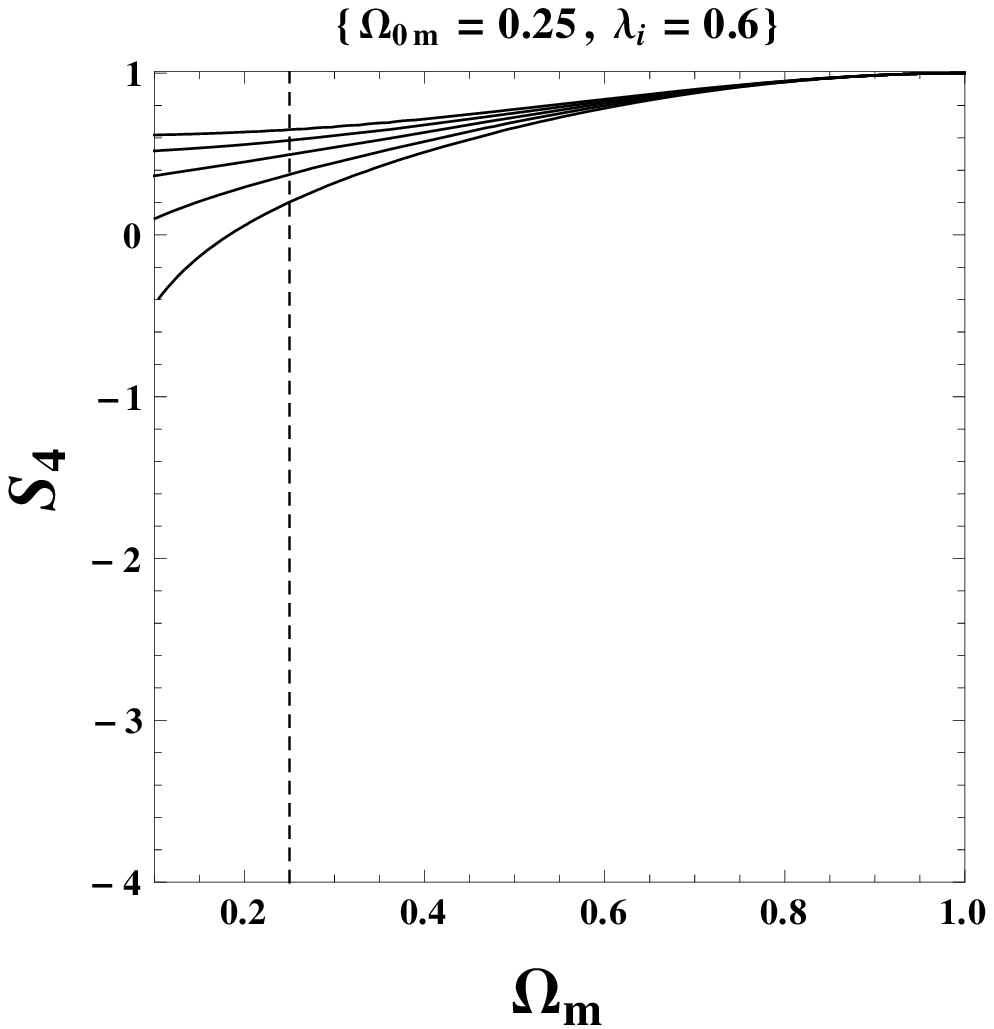}}
\end{tabular}
\caption{ Same as figure \ref{figs2} but for $S_{4}$.}
\label{figs4}
\end{center}
\end{figure}
\begin{figure}
\begin{center}
\begin{tabular}{c c }
{\includegraphics[width=2in,height=2in,angle=0]{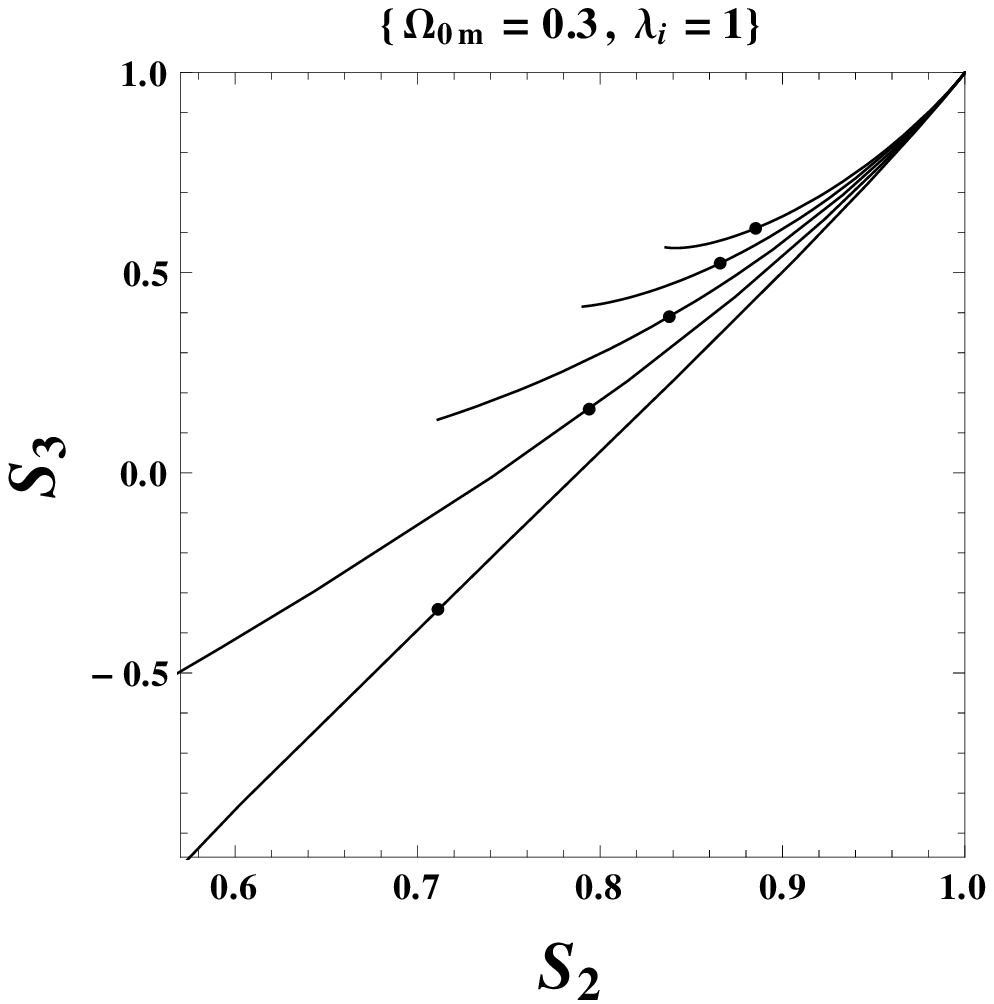}}&
{\includegraphics[width=2in,height=2in,angle=0]{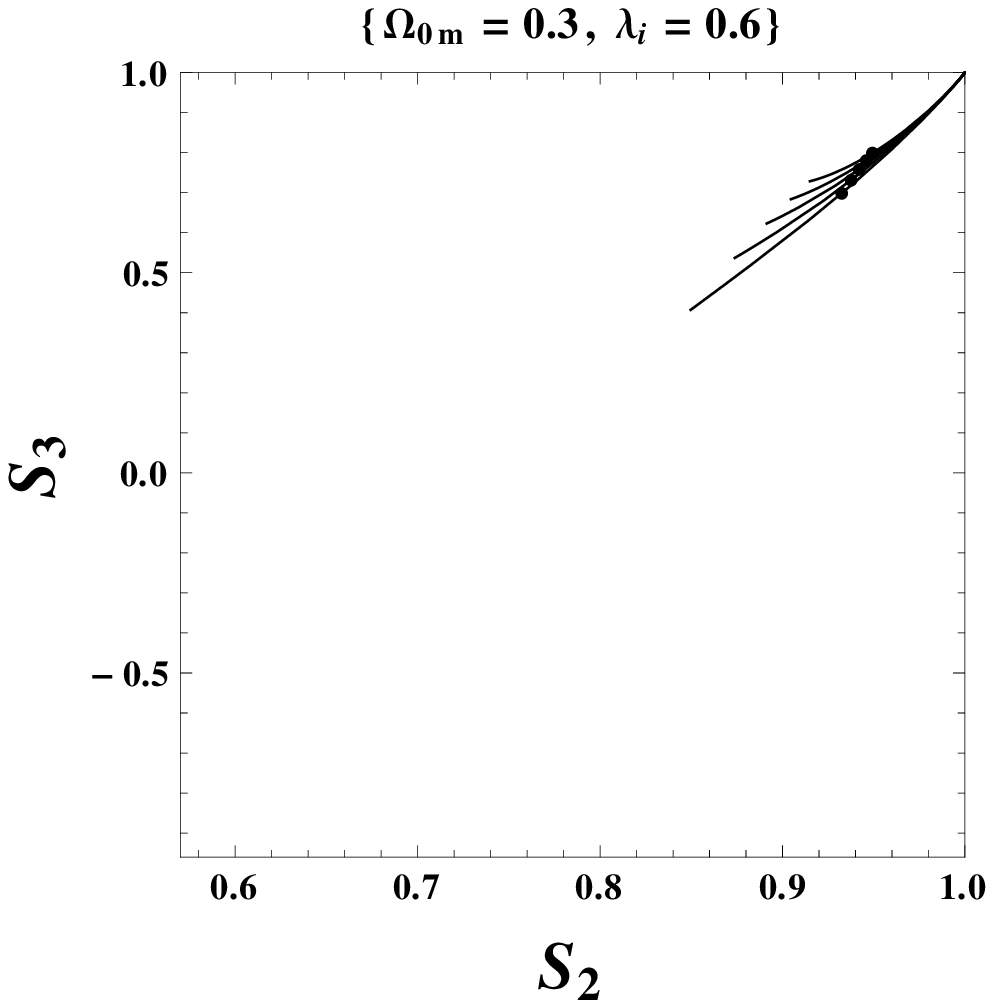}}
\\
{\includegraphics[width=2.1in,height=2.1in,angle=0]{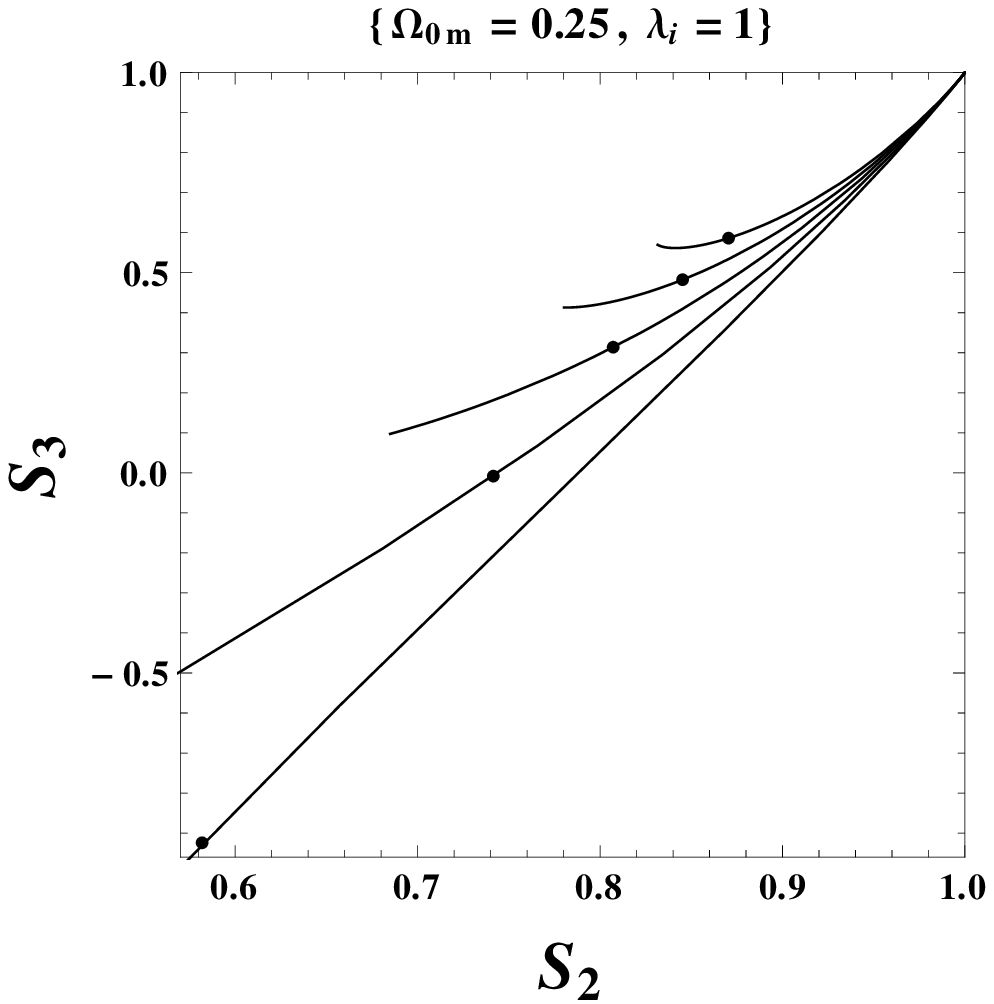}}&
{\includegraphics[width=2.1in,height=2.1in,angle=0]{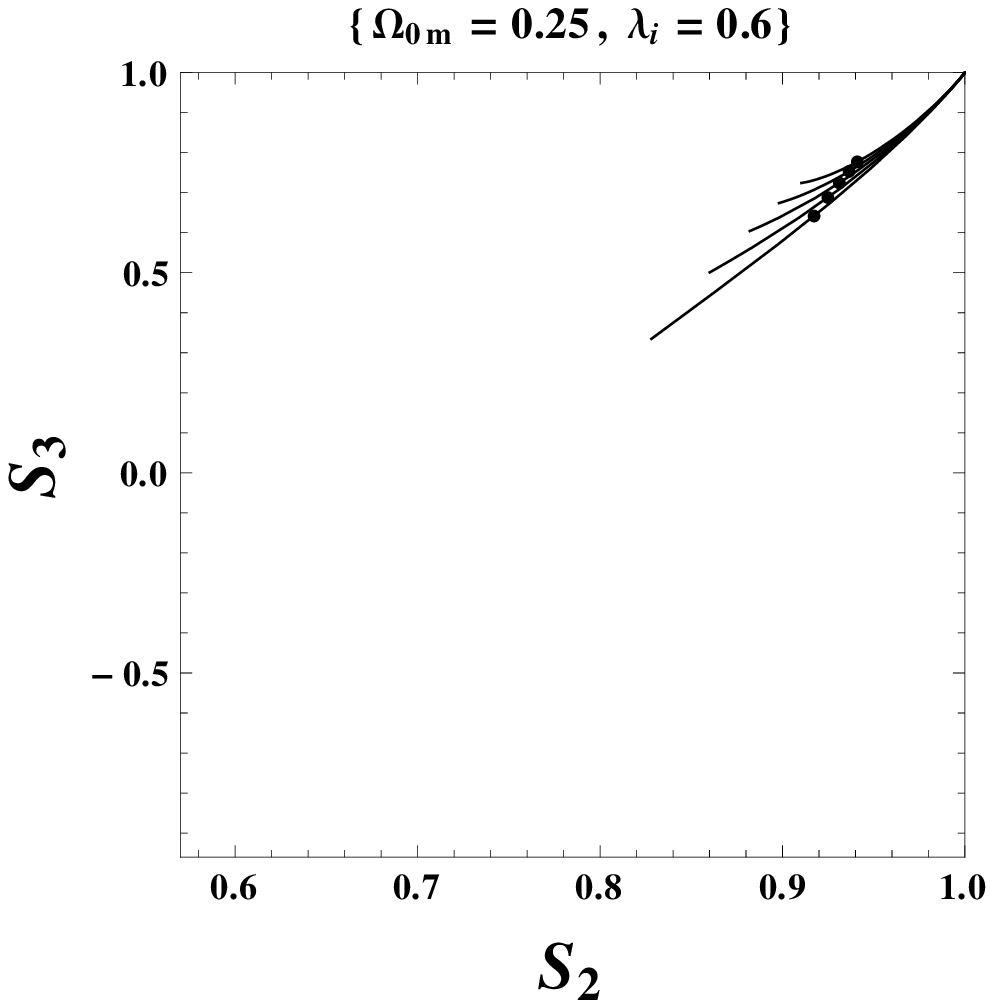}}
\end{tabular}
\caption{ This figure shows the evolution of $S_{3}$ versus $S_{2}$ for different potentials $V(\phi) \sim \phi, \phi^2, e^{\phi}, \phi^{-2}, \phi^{-1}$ from bottom to top and for different values of $\lambda_{i}$ and $\Omega_{0m}$. The black dots show the present epoch (z = 0).}
\label{figs3s2}
\end{center}
\end{figure}
where $\Omega_m = \omm (1+z)^3/h^2(z)$. It is notable to see that for $\Lambda$CDM, $S_n=1$ during the entire course of cosmic expansion. Now we use various combinations of $S_n$, to study the evolution of light mass galileon model with different potentials.

The initial value of $\lambda$ i.e $\lambda_i$ is an important parameter. It tells about the departure from the $\Lambda$CDM behaviour. Figure \ref{figwpi} shows that for smaller values of $\lambda_i$ ($\lambda_i = 0.1$) the models with different potentials can rarely be distinguished amongst themselves and from $\Lambda$CDM (w = -1). As $\lambda_i$ grows, all the models with different potentials start deviating from each other as well as from $\Lambda$CDM (w = -1). Furthermore, as we go for higher $\lambda_i$, the equation of state w$_\phi$ for linear potential has the largest departure from $\Lambda$CDM. In figure \ref{figs2}, we show the evolution of different potentials for different values of $\lambda_{i}$ and $\Omega_{0m}$ in $S_{2} - \Omega_{m}$ plane. As we have shown in the case of equation of state, here also the departure from the $\Lambda$CDM is small and large for smaller and larger values of $\lambda_i$ respectively. The linear potential shows the highest deviation from $\Lambda$CDM for $\lambda_i = 1$. The models with various potentials nearly degenerate for smaller values of $\lambda_i$ whereas for higher values of $\lambda_i$ the models are showing non degeneracy. Moreover, departure from the $\Lambda$CDM as well as among different potentials are larger for smaller values of $\Omega_{0m}$.

\begin{figure}
\begin{center}
\begin{tabular}{c c }
{\includegraphics[width=2.3in,height=2.1in,angle=0]{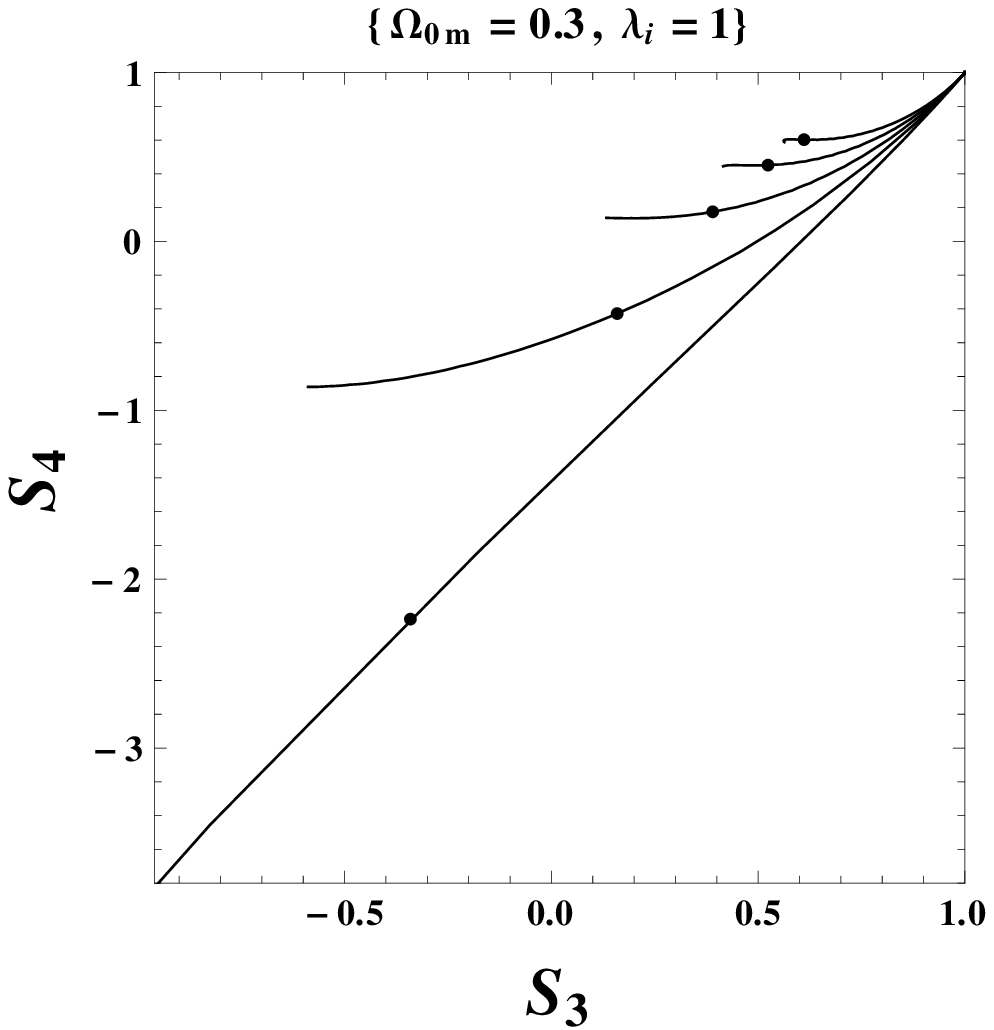}}&
{\includegraphics[width=2.3in,height=2.1in,angle=0]{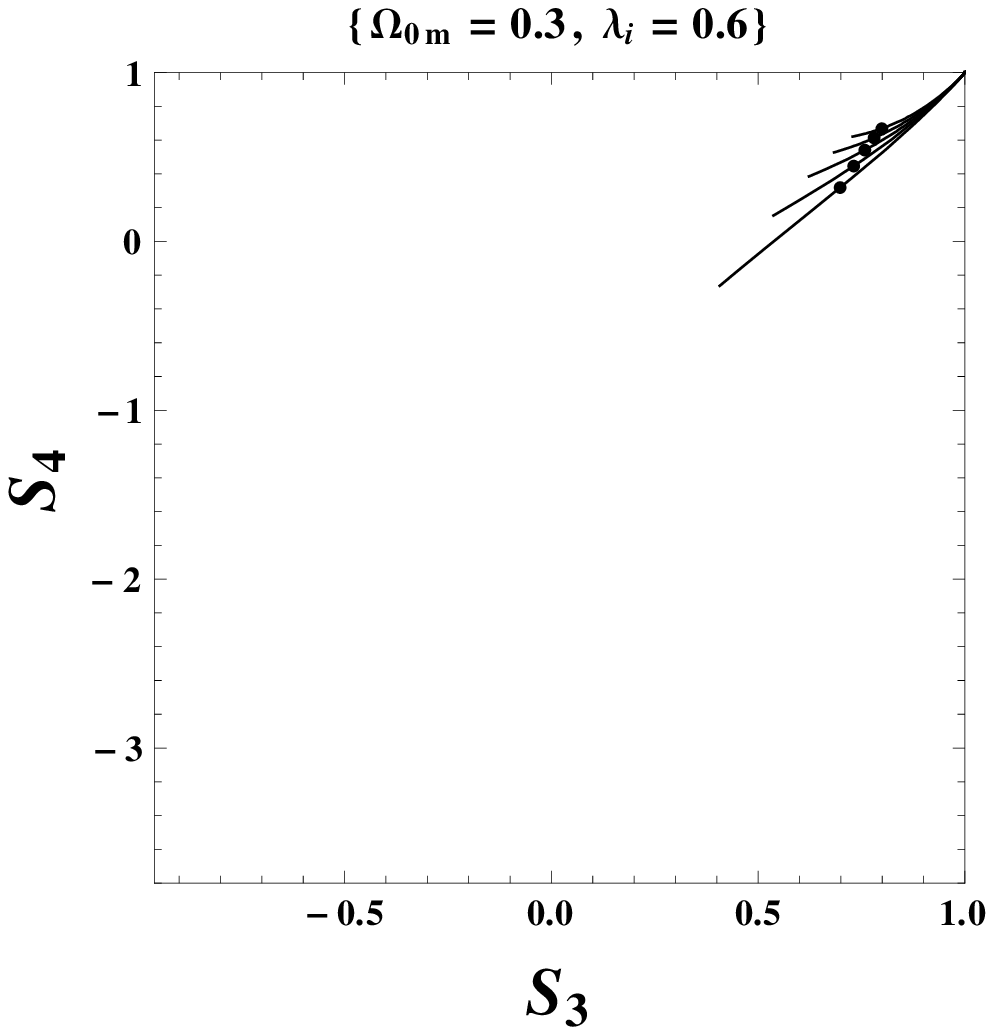}}
\\
{\includegraphics[width=2.3in,height=2.1in,angle=0]{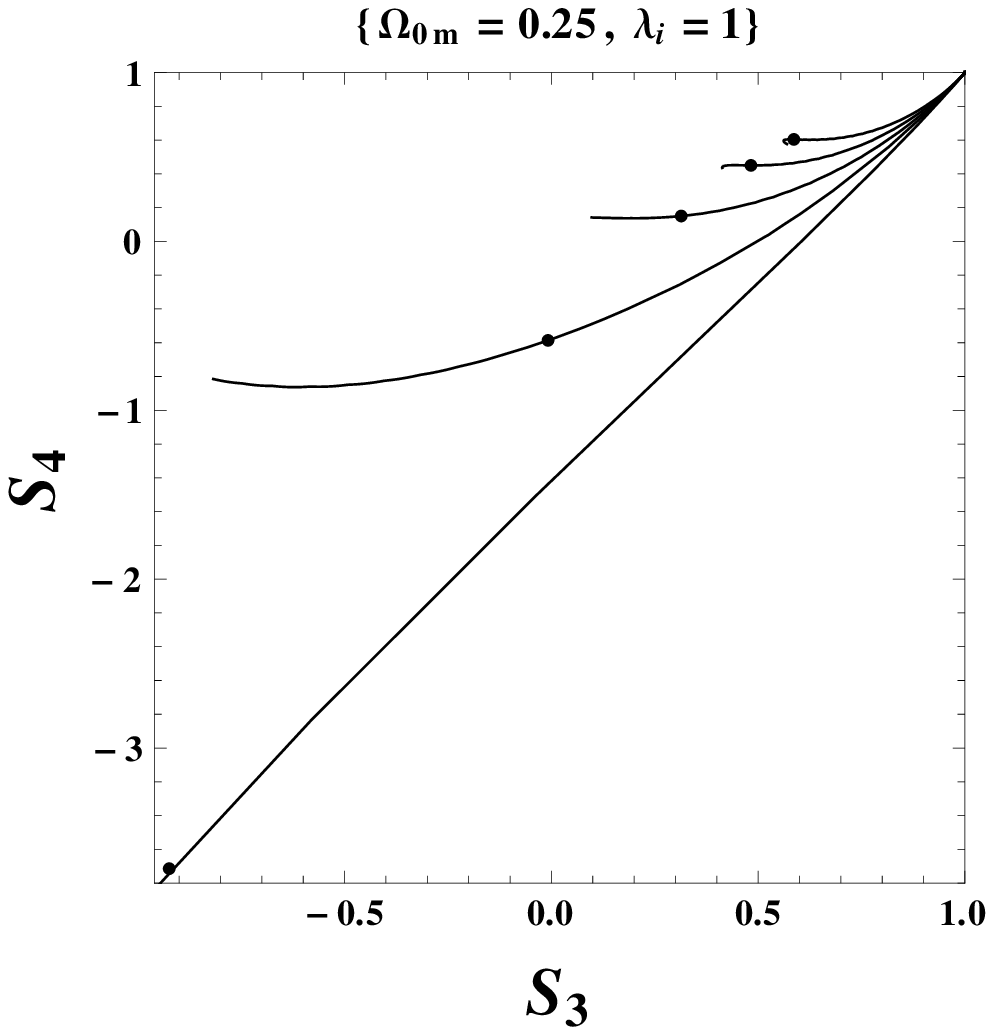}}&
{\includegraphics[width=2.3in,height=2.1in,angle=0]{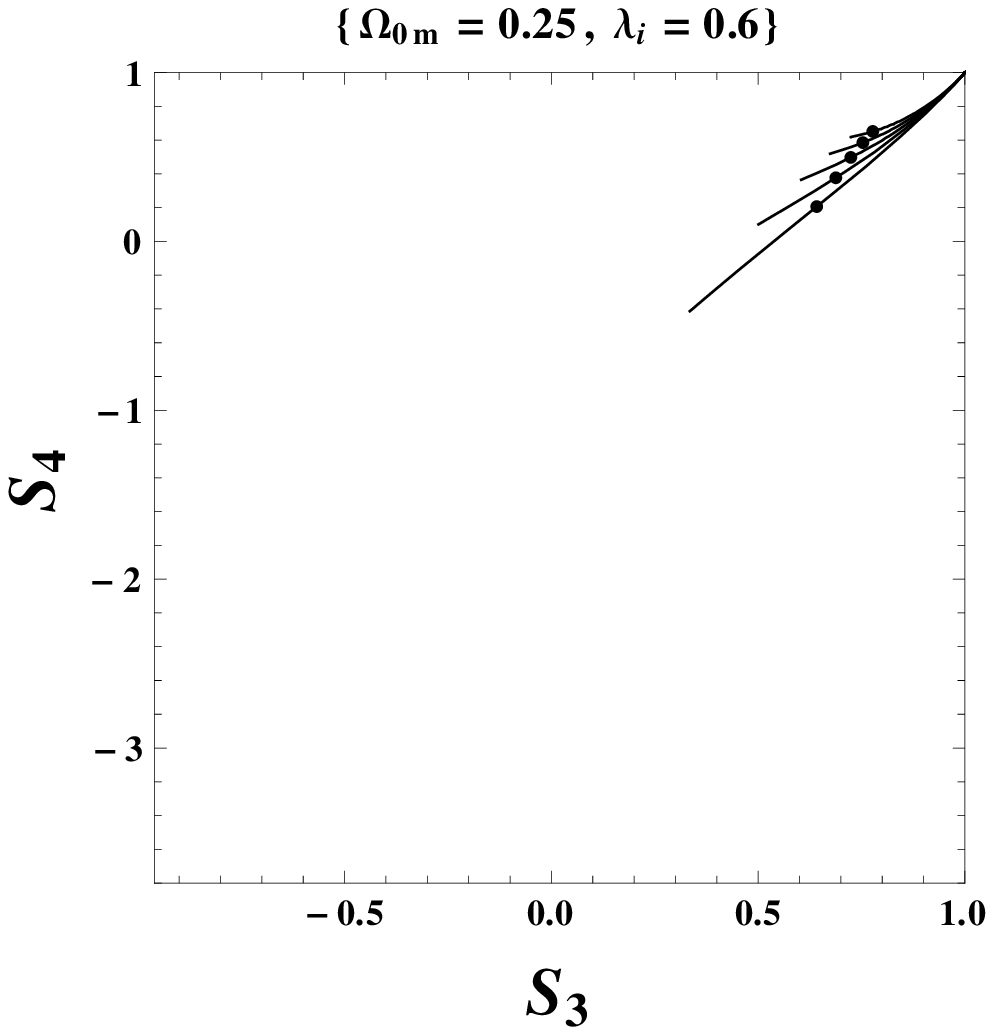}}
\end{tabular}
\caption{ Same as figure \ref{figs3s2} but for  $S_{4}$ versus $S_{3}$.}
\label{figs4s3}
\end{center}
\end{figure}

In figure \ref{figs3}, we show the evolution of models with different potentials for different values of $\lambda_{i}$ and $\Omega_{0m}$ in $S_{3} - \Omega_{m}$ plane. The departure from the $\Lambda$CDM as well as among different potentials are higher for smaller values of $\Omega_{0m}$. For smaller and larger values of $\lambda_{i}$ the models with different potentials nearly degenerate and non degenerate respectively. Next, we show the evolution of different potentials in the $S_{4} - \Omega_{m}$ plane in figure \ref{figs4}. Here too, the models with various potentials depart more for smaller $\Omega_{0m}$ and larger $\lambda_{i}$. In figures \ref{figs3s2} and \ref{figs4s3} we show the evolution of different potentials in the $S_{3} - S_{2}$ and  $S_{4} - S_{3}$ plane respectively. In these figures also the models with various potentials depart more for smaller and larger values of 
$\Omega_{0m}$ and $\lambda_{i}$ respectively. 
\section{$Om$ diagnostic}
\label{Om}
The $Om$, a geometrical diagnostic, is constructed from the hubble parameter and depends upon the first derivative of scale factor. It discriminates different dynamical dark energy models from $\Lambda$CDM with correct and incorrect values of the matter density. For $\Lambda$CDM model, $Om$ has same values at different redshifts. This implies that  non-evolving nature of $Om$ provides a null test for cosmological constant . 
The $Om$ for spatially flat Universe is defined as \cite{Om}:
\be
Om(x) \equiv \frac{H^2(x)/{H_0^2}-1}{x^3-1},~~~~ x = 1+z~.
\label{eq:om}
\ee
The hubble parameter for constant equation of state  is defined as,
\be
H^2(x) = H_0^2(\Omega_{0m} x^3 + (1-\Omega_{0m}) x^{3(1+w)}),
\label{eq:hubble}
\ee
Therefore,
\be
Om(x) = \Omega_{0m} + (1-\Omega_{0m})\frac{x^{3(1+w)}- 1}{x^3-1}~,
\label{Eq:omx}
\ee
from equation (\ref{Eq:omx}) we conclude that,

For $\Lambda$CDM $(w = -1)$, $Om(x) = \Omega_{0m}$, This implies that $Om$ has zero curvature.
For quintessence $(w > -1)$, $Om(x) < \Omega_{0m}$, This implies that $Om$ has negative curvature.
For phantom $(w < -1)$, $Om(x) > \Omega_{0m}$, This implies that $Om$ has positive curvature.

\begin{figure}
\begin{center}
\begin{tabular}{c c }
{\includegraphics[width=2.3in,height=2.1in,angle=0]{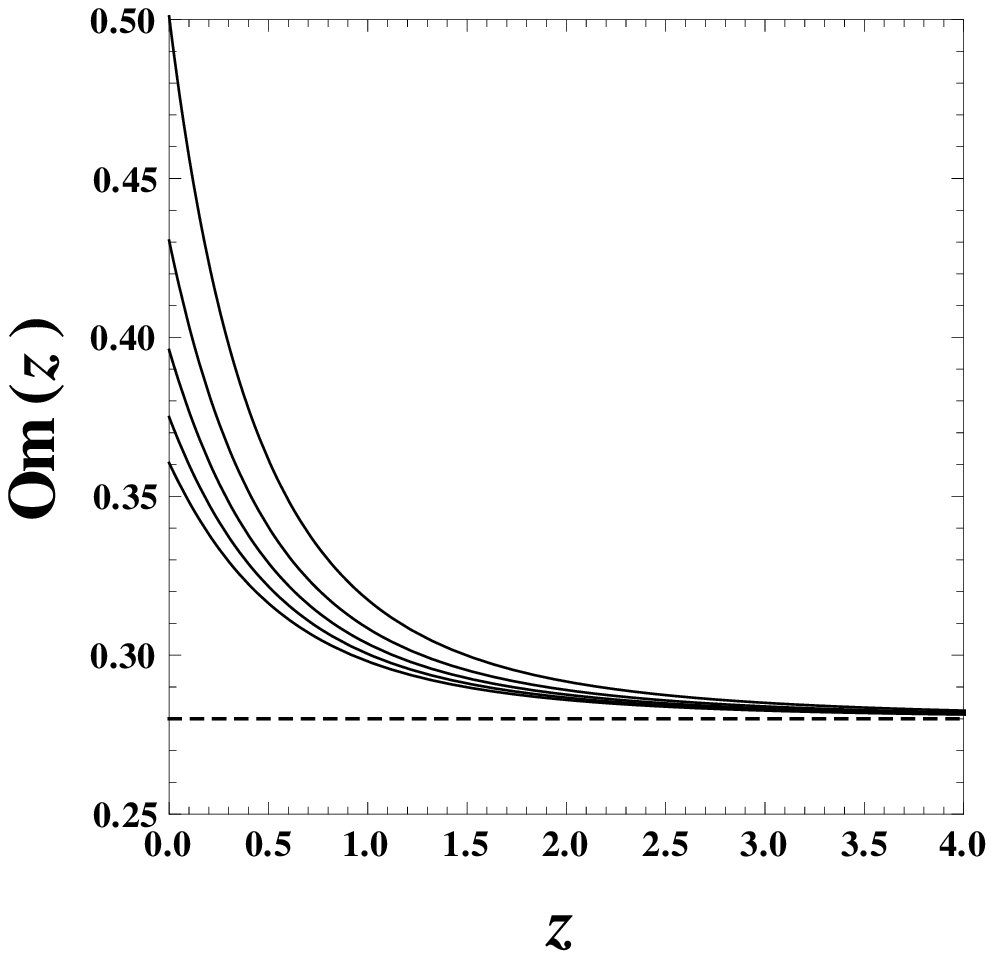}}&
{\includegraphics[width=2.3in,height=2.1in,angle=0]{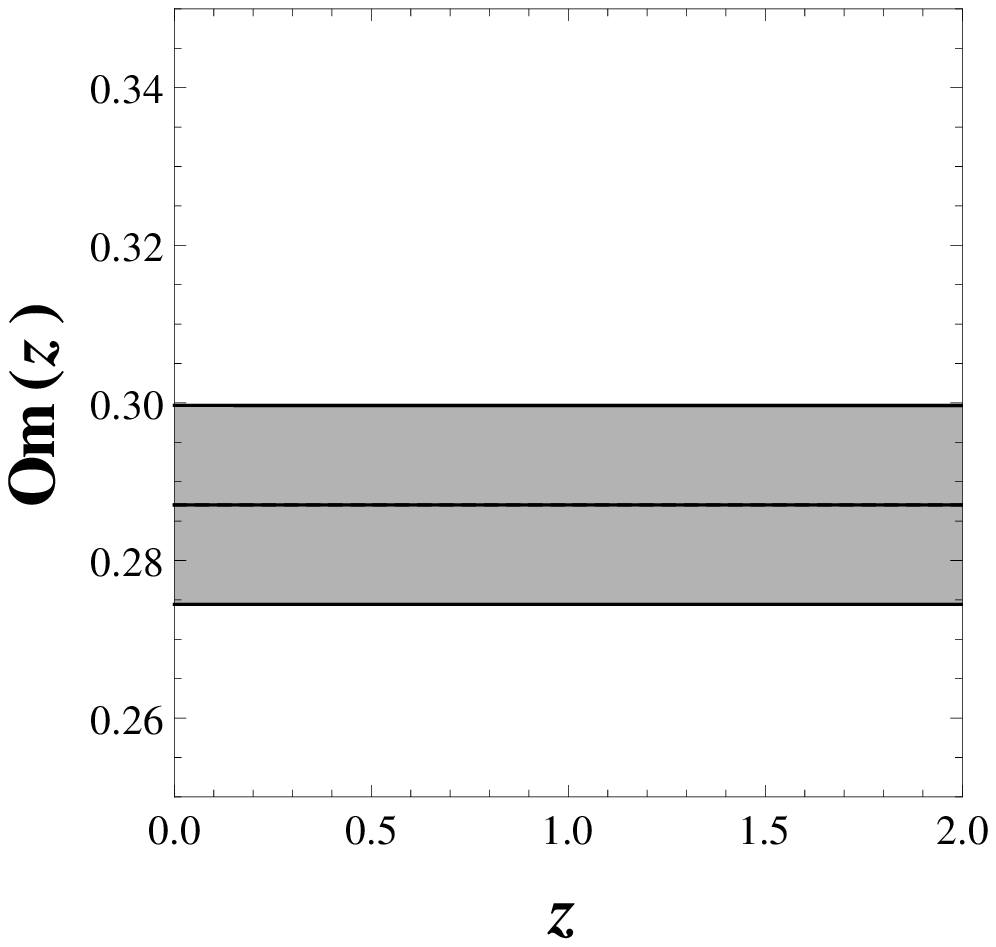}}
\end{tabular}
\caption{ This figure shows the evolution of $Om$ versus $z$. Left plot shows the $Om$ behaviour for the models with different potentials $V(\phi) \sim \phi, \phi^2, e^{\phi}, \phi^{-2}, \phi^{-1}$ from top to bottom with $\lambda_{i} = 1$ and $\Omega_{0m} = 0.28$. Right plot shows the $Om$ behaviour for the model with linear potential. The solid line shows the best fitted behaviour inside 1$\sigma$ confidance level. It looks same as $\Lambda$CDM due to the very small best fit value of $\lambda_{i} \approx 0.002505$. Here we use $\Omega_{0m} = 0.287057$.}
\label{figom}
\end{center}
\end{figure}

We, therefore, conclude that $Om(x) = \Omega_{0m}$ {\it iff} dark energy is a cosmological constant. It is interesting to see that $Om$ provides a null test of the $\Lambda$CDM hypothesis. In this section we want to show that $Om$ has negative curvature for quintessence dark energy models. The $Om$ behaviour for the models with different potentials is shown in the left plot of figure \ref{figom}, where $Om$ has negative curvature. In the right plot of figure \ref{figom} we show the best fitted behaviour inside 1$\sigma$ confidence level for the linear potential. The best fitted behaviour is constant and same as $\Lambda$CDM because the best fit value of the parameter $\lambda_{i} \approx 0.002505$ is very small. The best fitted behaviour of the models with other potentials  is same as $\Lambda$CDM due to the smaller best fit value of $\lambda_{i}$. This type of behaviour for equation of state is shown in figure \ref{figwpi} where for small values of $\lambda_{i}$ equation of state is nearly same as in case of $\Lambda$CDM. 
\begin{figure}
\begin{center}
\begin{tabular}{c c c }
{\includegraphics[width=2in,height=2in,angle=0]{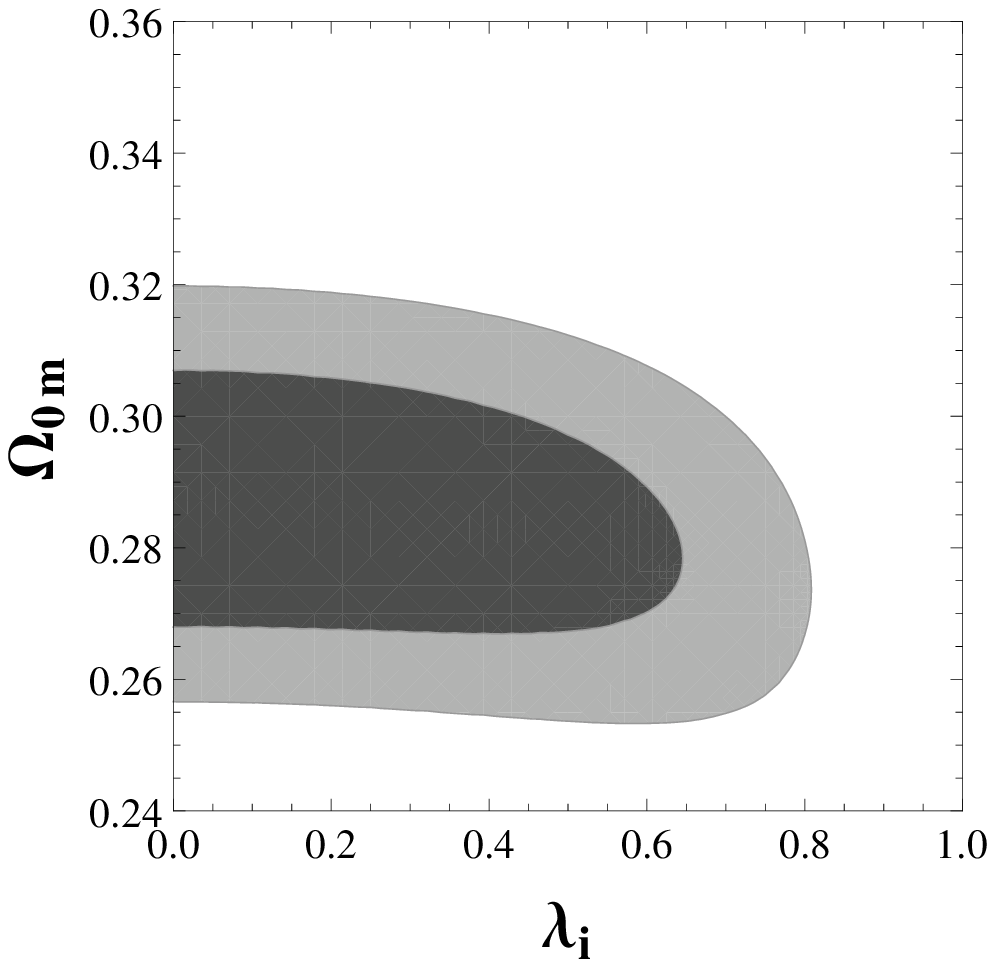}}&
{\includegraphics[width=2in,height=2in,angle=0]{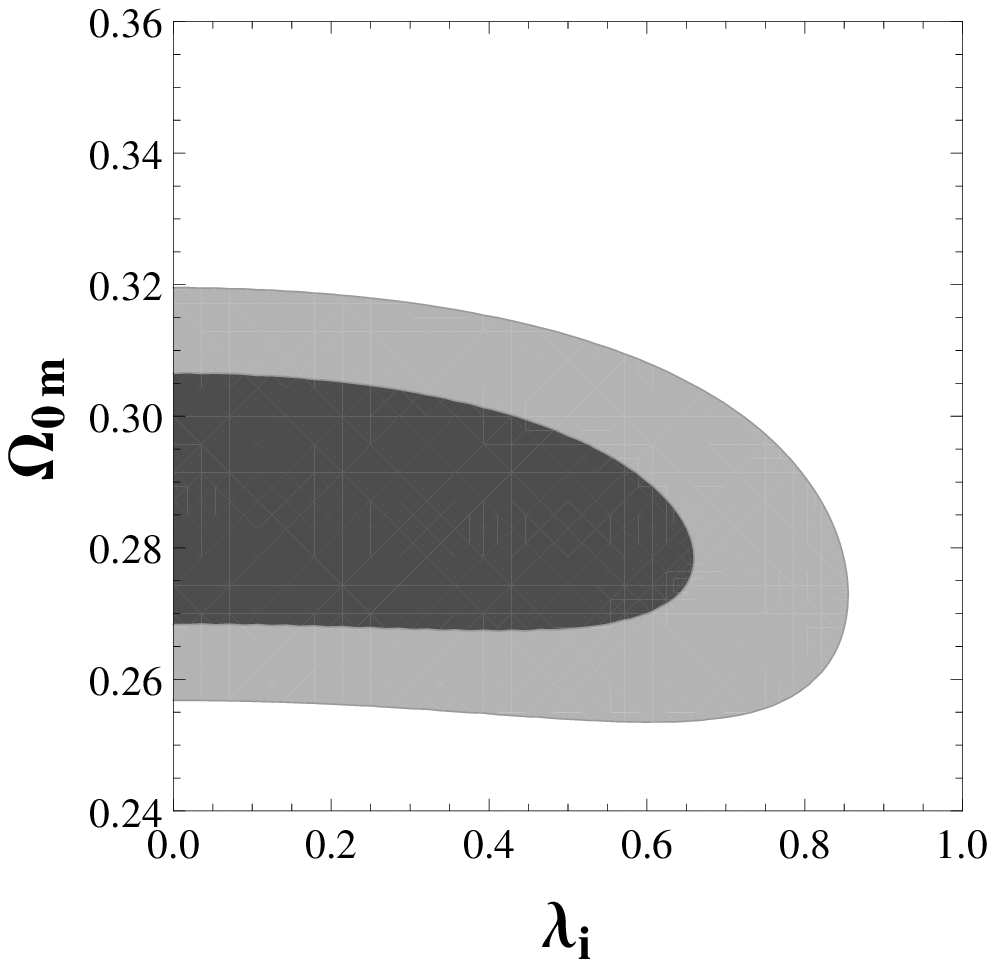}}&
{\includegraphics[width=2in,height=2in,angle=0]{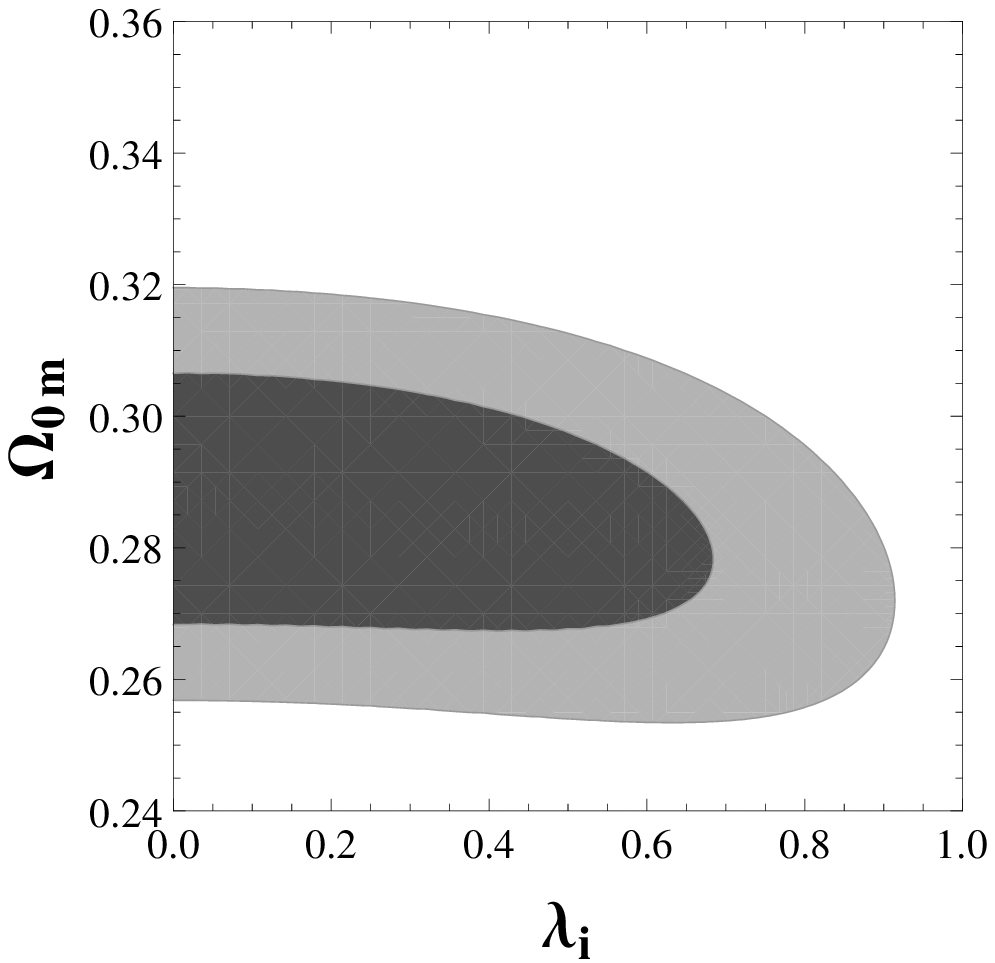}}
\end{tabular}
\caption{ This figure shows the 1$\sigma$ (dark shaded) and 2$\sigma$ (light shaded) likelihood contours in the $\lambda_{i} - \Omega_{0m}$ plane. Left, middle and right plots are for linear, quadratic and exponential potentials respectively.}
\label{figobs1}
\end{center}
\end{figure}
\begin{figure}
\begin{center}
\begin{tabular}{c c }
{\includegraphics[width=2in,height=2in,angle=0]{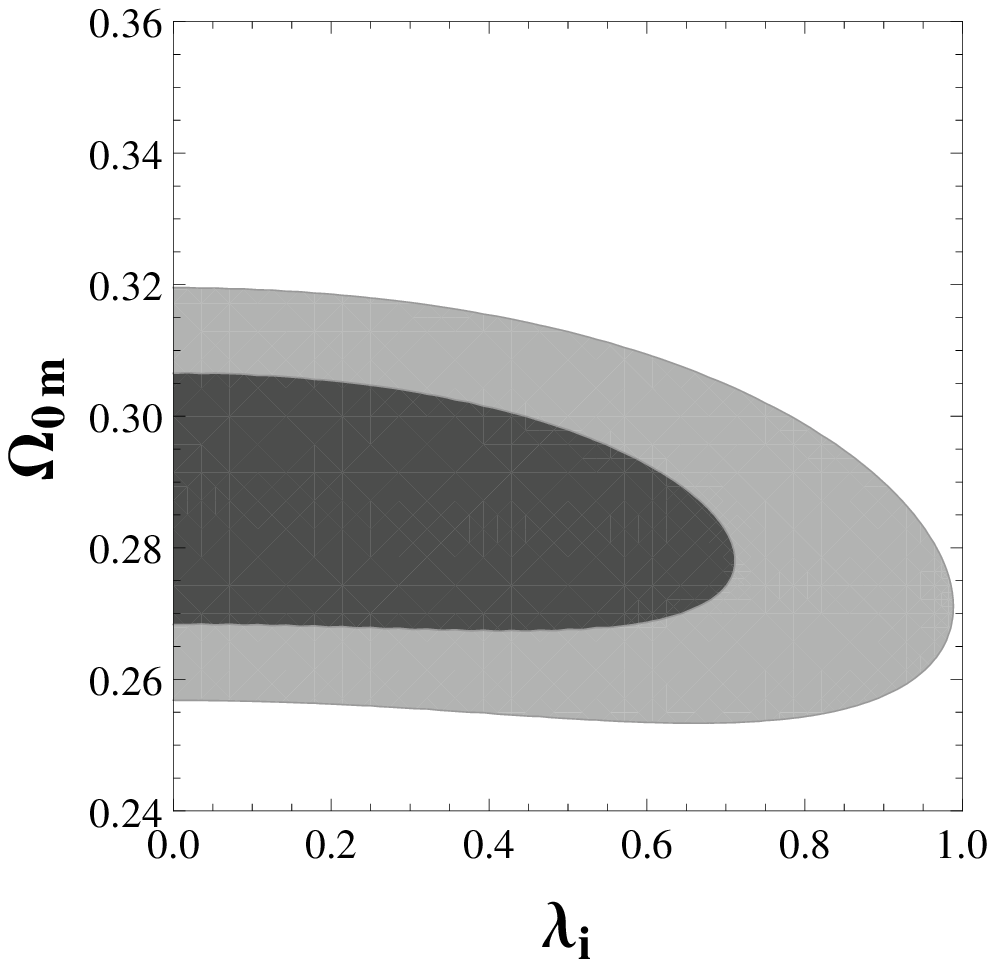}}&
{\includegraphics[width=2in,height=2in,angle=0]{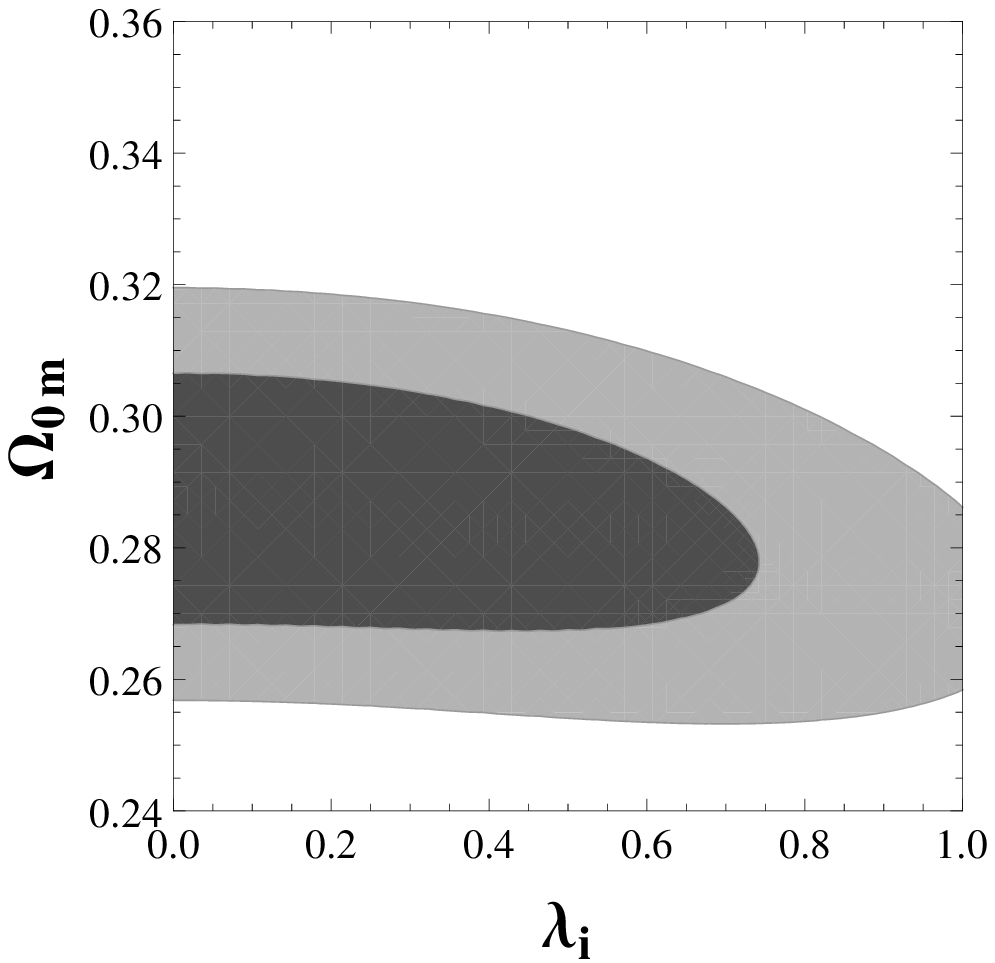}}
\end{tabular}
\caption{ Same as figure \ref{figobs1} but left one is for inverse square potential while right one is for inverse potential.}
\label{figobs2}
\end{center}
\end{figure}
\section{Observational constraints}
\label{obs}
We put observational constraints on the model parameters $\lambda_{i}$ and $\Omega_{0m}$ by applying latest observational data. We consider the supernova Type Ia observation which is one of the direct probes of the cosmic expansion. We use latest Union2.1 data compilation \cite{Suzuki:2011hu} consisting of 580 data points.

The observable quantity $\mu$ is the distance modulus which is defined as, $\mu=m - M=5 \log D_L+\mu_0$, where $m$ and $M$ are the apparent and absolute magnitudes of the supernovae,  $\mu_0$ is a nuisance parameter which is marginalized and $D_L$ is the luminosity distance defined  as $D_L(z)=(1+z)
\int_0^z\frac{dz'}{H(z')/H_0}$.

Next, we use latest 28 observational data points of hubble parameter at different redshifts compiled by Farroq et. al \cite{Farooq:2013hq}. We take $H_0$ from Planck 2013 results \cite{planck} to complete the data set. The values are shown in Table \ref{hubble}.

\begin{table}
\caption{$H(z)$ measurements (in units [$\mathrm{km\,s^{-1}Mpc^{-1}}$]) and their errors \cite{Farooq:2013hq}.}
\begin{center}
\label{hubble}
\begin{tabular}{cccc}
\hline\hline
~~$z$ & ~~~~$H(z)$ &~~~~ $\sigma_{H}$ & ~~ Reference\\
0.070&~~    69&~~~~~~~  19.6&~~ \cite{Zhang:2012mp}\\
0.100&~~    69&~~~~~~~  12&~~   \cite{Simon:2004tf}\\
0.120&~~    68.6&~~~~~~~    26.2&~~ \cite{Zhang:2012mp}\\
0.170&~~    83&~~~~~~~  8&~~    \cite{Simon:2004tf}\\
0.179&~~    75&~~~~~~~  4&~~    \cite{Moresco:2012by}\\
0.199&~~    75&~~~~~~~  5&~~    \cite{Moresco:2012by}\\
0.200&~~    72.9&~~~~~~~    29.6&~~ \cite{Zhang:2012mp}\\
0.270&~~    77&~~~~~~~  14&~~   \cite{Simon:2004tf}\\
0.280&~~    88.8&~~~~~~~    36.6&~~ \cite{Zhang:2012mp}\\
0.350&~~    76.3&~~~~~~~    5.6&~~  \cite{Chuang2012b}\\
0.352&~~    83&~~~~~~~  14&~~   \cite{Moresco:2012by}\\
0.400&~~    95&~~~~~~~  17&~~   \cite{Simon:2004tf}\\
0.440&~~    82.6&~~~~~~~    7.8&~~  \cite{Blake12}\\
0.480&~~    97&~~~~~~~  62&~~   \cite{Stern:2009ep}\\
0.593&~~    104&~~~~~~~ 13&~~   \cite{Moresco:2012by}\\
0.600&~~    87.9&~~~~~~~    6.1&~~  \cite{Blake12}\\
0.680&~~    92&~~~~~~~  8&~~    \cite{Moresco:2012by}\\
0.730&~~    97.3&~~~~~~~    7.0&~~  \cite{Blake12}\\
0.781&~~    105&~~~~~~~ 12&~~   \cite{Moresco:2012by}\\
0.875&~~    125&~~~~~~~ 17&~~   \cite{Moresco:2012by}\\
0.880&~~    90&~~~~~~~  40&~~   \cite{Stern:2009ep}\\
0.900&~~    117&~~~~~~~ 23&~~   \cite{Simon:2004tf}\\
1.037&~~    154&~~~~~~~ 20&~~   \cite{Moresco:2012by}\\
1.300&~~    168&~~~~~~~ 17&~~   \cite{Simon:2004tf}\\
1.430&~~    177&~~~~~~~ 18&~~   \cite{Simon:2004tf}\\
1.530&~~    140&~~~~~~~ 14&~~   \cite{Simon:2004tf}\\
1.750&~~    202&~~~~~~~ 40&~~   \cite{Simon:2004tf}\\
2.300&~~    224&~~~~~~~ 8&~~    \cite{Busca12}\\
\hline\hline
\end{tabular}
\end{center}
\end{table}

finally, we use BAO data of $\frac{d_A(z_\star)}{D_V(Z_{BAO})}$  \cite{Blake:2011en,Percival:2009xn,Beutler:2011hx,Jarosik:2010iu,Eisenstein:2005su,Giostri:2012ek},
where $d_A(z)=\int_0^z \frac{dz'}{H(z')}$ is the co-moving angular-diameter distance, $D_V(z)=\left(d_A(z)^2\frac{z}{H(z)}\right)^{\frac{1}{3}}$ is the dilation scale and $z_\star \approx 1091$ is the decoupling time. Data required for this analysis is shown in Table \ref{baodata}.
\\
\\
The $\chi_\mathrm{BAO}^2$ is described in reference \cite{Giostri:2012ek} and defined as,
\begin{equation}
 \chi_{\rm BAO}^2=X^t C^{-1} X\,,
\end{equation}
where,
\begin{equation}
X=\left( \begin{array}{c}
        \frac{d_A(z_\star)}{D_V(0.106)} - 30.95 \\
        \frac{d_A(z_\star)}{D_V(0.2)} - 17.55 \\
        \frac{d_A(z_\star)}{D_V(0.35)} - 10.11 \\
        \frac{d_A(z_\star)}{D_V(0.44)} - 8.44 \\
        \frac{d_A(z_\star)}{D_V(0.6)} - 6.69 \\
        \frac{d_A(z_\star)}{D_V(0.73)} - 5.45
        \end{array} \right)\,,
\end{equation}
and the inverse covariance matrix,
\begin{align}
C^{-1}=\left(
\begin{array}{cccccc}
 0.48435 & -0.101383 & -0.164945 & -0.0305703 & -0.097874 & -0.106738 \\
 -0.101383 & 3.2882 & -2.45497 & -0.0787898 & -0.252254 & -0.2751 \\
 -0.164945 & -2.45499 & 9.55916 & -0.128187 & -0.410404 & -0.447574 \\
 -0.0305703 & -0.0787898 & -0.128187 & 2.78728 & -2.75632 & 1.16437 \\
 -0.097874 & -0.252254 & -0.410404 & -2.75632 & 14.9245 & -7.32441 \\
 -0.106738 & -0.2751 & -0.447574 & 1.16437 & -7.32441 & 14.5022
\end{array}
\right)\,.
\end{align}
\begin{table*}[!]
\caption{Values of $\frac{d_A(z_\star)}{D_V(Z_{BAO})}$ for different values
of $z_{BAO}$.}
\begin{center}
\resizebox{\textwidth}{!}{%
\begin{tabular}{c||cccccc}
\hline\hline
\\
 $z_{BAO}$  & 0.106~~  & 0.2~~ & 0.35~~ & 0.44~~ & 0.6~~ & 0.73\\
\hline
 $\frac{d_A(z_\star)}{D_V(Z_{BAO})}$ &  $30.95 \pm 1.46$~~ & $17.55 \pm 0.60$~~
& $10.11 \pm 0.37$ ~~& $8.44 \pm 0.67$~~ & $6.69 \pm 0.33$~~ & $5.45 \pm 0.31$
\\
\\
\hline\hline
\end{tabular}}
\label{baodata}
\end{center}
\end{table*}
\begin{table}
\caption{Best fit values of the model parameters for different potentials.}
\begin{center}
\label{bestfit}
\begin{tabular}{ccc}
\hline\hline
~~Potentials & ~~~~$\lambda_i$ &~~~~ $\Omega_{0m}$\\
\hline
\\
$V(\phi)\propto \phi$&~~    0.002505&~~~~~~~  0.287057\\
\\
$V(\phi)\propto \phi^2$&~~    0.002165&~~~~~~~  do\\
\\
$V(\phi)\propto e^\phi$&~~    0.001884&~~~~~~~ do\\
\\
$V(\phi)\propto \phi^{-2}$&~~    0.003233&~~~~~~~ do\\
\\
$V(\phi)\propto \phi^{-1}$&~~    0.002964&~~~~~~~ do\\
\\
\hline\hline
\end{tabular}
\end{center}
\end{table}
The results are shown in figures \ref{figobs1} and \ref{figobs2} where we show 1$\sigma$ (dark shaded) and 2$\sigma$ (light shaded) likelihood contours in the $\lambda_{i} - \Omega_{0m}$ plane. The right plot of figure \ref{figobs2} shows that inverse potential has highest allowed deviation from the $\Lambda$CDM behaviour. The best fit values of the model parameters are shown in Table \ref{bestfit}.
\section{Conclusions}
\label{conc}
In this paper, we restrict ourselves to the lowest order galileon lagrangian ${\cal L}_3$ but add a general potential term $V(\phi)$ to the Lagrangian  and explore the late time cosmological evolution of light mass galileon with different choices for $V(\phi)$. We acquire that the $\phi$ field is initially frozen due to large hubble friction and acts as a cosmological constant. We do not acquire slow roll conditions for the potentials under consideration thereby $\lambda_{i}$ is a free parameter in the model. The deviation from $w = -1$ ($\Lambda$CDM) depends upon the value of $\lambda_{i}$. For smaller values of $\lambda_{i}$, the departure is  small and all potentials behave like  cosmological constant throughout. As the value of $\lambda_{i}$ grows, the evolution begins departing from $w = -1$ ($\Lambda$CDM). By applying statefinder hierarchy, we discuss degeneracies for the various potentials. It is found  that $S_3$ is best suited for removing the degeneracy amongst the models we considered in case of $\Omega_{0m}\simeq 0.25,\lambda_{i}\simeq 1$. However, the same lies out side 1$\sigma$ bound. We should admit that shift symmetry in the Minkowski background breaks by adding a potential. However, since the mass of galileon is of the order of $H_0$, the effect of symmetry breaking is mild.

We also use $Om$ diagnostic to show that $Om$ has negative slope for the models having equation of state $w > -1$, and this is shown in the left plot of figure \ref{figom} for $\lambda_{i}=1$. The right plot of figure \ref{figom} shows that $Om$ acts like a cosmological constant due to the small best fit value of the parameter $\lambda_{i}$. We used SN+Hubble+BAO data to constraint the model parameters.
\section*{Acknowledgement}
We thank M. Sami for his useful comments and suggestions. MS thanks to Sumit Kumar and M. W. Hossain for fruitful discussions.

\end{document}